\documentclass[9pt,twocolumn,twoside]{osajnl}
\usepackage{textcomp}

\def\beq{\begin{equation}}
\def\eeq{\end{equation}}

{\catcode`\|=\active 
  \gdef\Braket#1{\left<\mathcode`\|"8000\let|\bravert {#1}\right>}}
\def\bravert{\egroup\,\vrule\,\bgroup}

\def\itpm{\tilde{i}^\pm} 
\def\ptpm{\tilde{p}^\pm} 
 
\def\Ptpm{\tilde{P}^\pm} 

\journal{josab} % Choose journal (ao, aop, josaa, josab, ol)

\setboolean{shortarticle}{false} 
\title{Thermal Modeling, Heat Mitigation, and Radiative Cooling for Double-Clad Fiber Amplifiers}

\author[1,2]{Esmaeil Mobini}
\author[1,2]{Mostafa Peysokhan}
\author[1,2]{Behnam Abaie}
\author[1,2,*]{Arash Mafi}

\affil[1]{Department of Physics \& Astronomy, University of New Mexico, Albuquerque, NM 87131, USA}
\affil[2]{Center for High Technology Materials, University of New Mexico, Albuquerque, NM 87106, USA}

\affil[*]{Corresponding author: mafi@unm.edu}

\ociscodes{}

\begin{abstract}
We report a detailed formalism aimed at the thermal modeling and heat mitigation in high-power double-clad fiber amplifiers.
Closed form analytical formulas are developed that take into account the spatial profile of the amplified signal and pump 
in the double-clad geometry, the presence of the amplified spontaneous emission, and the possibility of radiative cooling due
to anti-Stokes fluorescence emission. The formalism is applied to a high-power Yb-doped silica fiber amplifier. The contributions
to the heat-load from the pump-signal quantum defect, as well as the pump and signal parasitic absorptions are compared 
to the radiative cooling. It is shown that for realistic cases, the local heat generation in kiloWatt-class fiber amplifiers is 
either dominated by the quantum defect or the parasitic absorption depending on the pump wavelength. In conventional designs,
radiative cooling can be substantial only in properly designed amplifiers, when the pump power is tens of watts or lower, unless
the parasitic absorption is reduced compared to the commonly reported values in the literature. 
We also explore the impact of the non-ideal quantum efficiency of the gain material. The developed formalism can be used to 
design fiber amplifiers and lasers for optimal heat mitigation, especially due to radiative cooling.
\end{abstract}

\setboolean{displaycopyright}{true}

\begin{document}

\maketitle

%%%%%%%%%%%%%%%%%%%%%%%%%%%%%%%%%%%%%%%%%%%%%%%%%%%%%%
%%%%%%%%%%%%%%%%%%%%%%%%%%%%%%%%%%%%%%%%%%%%%%%%%%%%%%
\section{Introduction}
%%%%%%%%%%%%%%%%%%%%%%%%%%%%%%%%%%%%%%%%%%%%%%%%%%%%%%
%%%%%%%%%%%%%%%%%%%%%%%%%%%%%%%%%%%%%%%%%%%%%%%%%%%%%%
High-power rare-earth-doped double-clad fiber lasers and amplifiers have proven to be reliable and versatile for many industrial and directed energy applications, 
enjoy excellent power efficiencies, and provide near diffraction-limited beam qualities~\cite{Richardson}. In the path to obtain ever increasing output powers from
fiber lasers and amplifiers, the geometrical form factor of the optical fiber plays an important role for heat mitigation. However, recent advances in
power scaling of fiber lasers and amplifiers are hindered by the thermally induced transverse mode instability, which deteriorates the 
output laser beam quality~\cite{brown2001thermal,zenteno1993high,ward2012origin,Dawson:08,jauregui2012physical}.
The obvious solution is to use ever more efficient heat removal schemes. 

Radiative cooling has been suggested as a potentially attractive method to mitigate the heat generation in fiber lasers and amplifiers~
\cite{bowman1999lasers,bowman2010minimizing,BowmanOptEng,yang2018radiation}.
The rare-earth-doped optical fiber is pumped at a wavelength $\lambda_p$, which is higher than the mean fluorescence wavelength 
$\lambda_f$ of the active ions; therefore, the anti-Stokes fluorescence removes some of the excess heat. Ideally, it may be possible
to fully compensate the heat generation in the gain medium by radiative cooling and operate as a radiation-balanced laser or amplifier
~\cite{S.R.Bowmancleo,yang2018radiation,nemova2009athermal,nemova2011radiation,Peysokhan:17}.
However, the pump absorption cross section is typically considerably lower than its peak value if $\lambda_p$ is chosen to be sufficiently
larger than $\lambda_f$ for efficient radiative cooling. Therefore, one naturally expects this to impact the design of the laser oscillator
or amplifier for proper operation.
 
In this paper, we explore the thermal modeling and heat mitigation including radiative cooling of high-power double-clad fiber amplifiers.
We present a detailed formalism that takes into account the spatial profile of the amplified signal and the pump in the double-clad geometry, 
the presence of the amplified spontaneous emission (ASE), and the possibility of radiative cooling~\cite{giles1991modeling,knall2018model,DigonnetJLT}.
 The formalism is then used for a specific case study of a high-power silica fiber amplifier. The intention is not to find an optimum design; rather, to explore the 
impact of various design parameters that affect the heat load of a fiber amplifier. The heat is generated due to the pump-signal
quantum defect, as well as the parasitic absorptions of the pump and signal. The former can generally be reduced by using a pump wavelength 
that is close to the wavelength of the amplified signal, while the latter is broadly a fiber material design issue. However, other 
design parameters of the amplifier can have considerable indirect impact on the heat load contribution from either source. 

We already mentioned that the wavelength  of the pump (and the amplified signal) must be larger than $\lambda_f$ for radiative cooling. 
For efficient radiative cooling in fiber lasers and amplifiers, the cooling material--in this case the doped fiber core--must ideally be 
cooling grade and have a high quantum 
efficiency~\cite{epstein1995observation,epstein2010optical,seletskiy2016laser,melgaard2016solid,nguyen2013towards,seletskiy2010laser,bowman2000new}. 
This puts a stringent constraint on the type and purity of the host glass. For example, the quantum efficiency of ytterbium-doped fluorozirconate 
glass (Yb:ZBLAN) has been measured to be as high as 0.995~\cite{DenisZBLAN,gosnell1998laser,gosnell1999laser,Peysokhan:18}, making it a desirable 
material for radiative cooling~\cite{epstein1995observation,gosnell1998laser,gosnell1999laser,thiede2005cooling,edwards1998development}.
In this paper, most of our numerical simulations are carried out for perfect quantum efficiency ($\eta_q\,=\,1$) in Yb-doped silica fibers.
The reason for choosing silica as the host material is to explore realistic designs related to previously published experimental work 
on high-power Yb-doped fiber amplifiers, all of which are in silica-based fibers. 

To the best of our knowledge, there is no published work on radiative cooling of Yb-doped silica glass. However, 
we think that there is no fundamental reason why high-purity Yb-doped silica 
cannot have very high quantum efficiency~\cite{pask1995ytterbium,digonnet2001rare,faure2007improvement,Mobini:18}. 
Our choice of perfect quantum efficiency is merely intended to show the upper limit of what is 
achievable by radiative cooling, while realistic designs and systematic studies for the impact of the cooling efficiency 
are left for future studies. 
%for future researches
The final simulation related to Fig.~\ref{Fig:eta-80-heat-versus-position-1030nm-20Wpump-50mlength} in this paper is carried out
for a value of $\eta_q\,=\,0.8$ to illustrate that heat mitigation due to anti-Stokes fluorescence emission in fiber amplifiers and 
lasers is still viable in less-than-ideal materials with $\eta_q\,<\,1$. 
%%%%%%%%%%%%%%%%%%%%%%%%%%%%%%%%%%%%%%%%%%%%%%%%%%%%%%
%%%%%%%%%%%%%%%%%%%%%%%%%%%%%%%%%%%%%%%%%%%%%%%%%%%%%%
\section{Basic formalism}
%%%%%%%%%%%%%%%%%%%%%%%%%%%%%%%%%%%%%%%%%%%%%%%%%%%%%%
%%%%%%%%%%%%%%%%%%%%%%%%%%%%%%%%%%%%%%%%%%%%%%%%%%%%%%
The optical fiber considered here has a step-index core of radius $a$, which is uniformly doped with Yb ions at a density $N_0$. 
The pump propagates in a double-clad geometry of radius $b$ and is assumed to be uniformly distributed over the transverse cross
section of the heavily multimode inner cladding. The core is assumed to be quasi-single-mode~\cite{beier2017single}; therefore, 
the signal (laser and ASE) propagate with a nearly Gaussian intensity profile described by 
%%%%%%%
\begin{align}
g_w(r) =\exp(-2\,r^2/w^2),
\end{align}
%%%%%%%
where $r$ is the radial coordinate and $w$ is the field radius. This Gaussian approximation is reliable for our purposes and
$w$ can be determined from an analytic approximation accurate to within 1\% for the core $V$ parameter in the range
$1.2 < V < 2.4$ given by~\cite{Marcuse:78,agrawal2012fiber}
%%%%%%%
\begin{align}
w/a\approx 0.65 + 1.619\,V^{-3/2} + 2.879\, V^{-6}.
\end{align}
%%%%%%%
The total ion concentration $N_t$ is given by
%%%%%%%
\begin{align}
N_t(r) = N_0\,\Theta_a(r), \qquad 
  \Theta_a(r) = 
  \begin{cases}
    1 & \text{for\ } 0 \leq r \leq a \\
    0 & \text{for\ } a < r 
  \end{cases}.
\end{align}
%%%%%%%

The rate equation for the upper manifold population density $N_2$ is given by 
\begin{align}
%%%%%%%
\nonumber
\dfrac{dN_2}{dt}
&=\dfrac{I^+_p+I^-_p}{h\nu_p}\Big[\sigma^a_p\,N_t-(\sigma^a_p+\sigma^e_p)\,N_2\Big]\\
&+\sum\limits^n_{j=1}\dfrac{I^+_j+I^-_j}{h\nu_j}\Big[\sigma^a_jN_t-(\sigma^a_j+\sigma^e_j)N_2\Big]
-\dfrac{N_2}{\tau_f}.
\end{align}
%%%%%%%
The pump frequency (wavelength) is $\nu_p$ ($\lambda_p$). The signal spectrum is sliced into $n$
adjacent segments, where $\delta\lambda$ is the bandwidth for each segment. The signal frequencies and wavelengths 
are $\nu_j$ ($\lambda_j$), where $j=1,\cdots,n$. The emission and absorption cross sections are defined as
%%%%%%%
\begin{align}
\sigma^e_j=\sigma^e(\lambda_j),
\quad
\sigma^a_j=\sigma^a(\lambda_j),
\quad
j=1,\cdots,n, \ {\rm or}\ j=p.
\end{align}
%%%%%%%
The pump and signal local intensities are defined as $I^\pm_j$ for $j=p$ and $j=1,\cdots,n$, respectively. The 
$\pm$ superscripts indicate forward and backward propagation. Note that $I^\pm_j$ and $N_2$ are functions of both
radial ($r$) and longitudinal ($z$) coordinates. The upper manifold lifetime is $\tau_f=1/(\tau^{-1}_{\rm r}+\tau^{-1}_{\rm nr})$,
where $\tau_{\rm r}$ is the radiative lifetime and $\tau_{\rm nr}$ is the non-radiative lifetime.
The (internal) quantum efficiency is defined as
%%%%%%%
\begin{align}
\eta_q\,:=\,\dfrac{\tau_{\rm nr}}{\tau_{\rm r}+\tau_{\rm nr}}.
\label{Eq:etaq}
\end{align}
%%%%%%%
Throughout this paper, we assume that due to small size of the fiber core, the external quantum efficiency is very
close to the internal quantum efficiency~\cite{epstein1995observation,ruan2006enhanced}.

The steady-state condition for the upper manifold population ($dN_2/dt=0$) results in
%%%%%%%
\begin{align}
n_2=\dfrac{\beta_p\,(i^+_p+i^-_p)+\sum\limits^n_{j=1}\beta_j\,(i^+_j+i^-_j)}{1+(i^+_p+i^-_p)+\sum\limits^n_{j=1}(i^+_j+i^-_j)},
\end{align}
%%%%%%%
Here, $n_2(r,z)=N_2(r,z)/N_0$ and we have used the following definitions:
%%%%%%%
\begin{align}
\beta_j&=\dfrac{\sigma^a_j}{\sigma^a_j+\sigma^e_j},\quad &&j=1,\cdots,n, \ {\rm or}\ j=p,\\
i^\pm_j&=\dfrac{I^\pm_j}{I^{\rm sat}_j},\quad &&I^{\rm sat}_j=\dfrac{h\nu_j\,\beta_j}{\tau_f\,\sigma^a_j}.
\end{align}
%%%%%%%
Later in this paper, we will frequently encounter sums of the forward and backward propagating pump and signal intensities. In order to simplify, 
we identify such sums with a tilde overline:   
%%%%%%%
\begin{align}
\itpm_j=i^+_j+i^-_j
,\quad
j=1,\cdots,n, \ {\rm or}\ j=p.
\end{align}
%%%%%%%
Another definition that will prove useful later is
%%%%%%%
\begin{align}
\gamma_j=1-\beta_j/\beta_p,
\quad
j=1,\cdots,n.
\end{align}
%%%%%%%
%%%%%%%%%%%%%%%%%%%%%%%%%%%%%%%%%%%%%%%%%%%%%%%%%%%%%%
%%%%%%%%%%%%%%%%%%%%%%%%%%%%%%%%%%%%%%%%%%%%%%%%%%%%%%
\subsection{Pump propagation}
\label{sec:pumpprop}
%%%%%%%%%%%%%%%%%%%%%%%%%%%%%%%%%%%%%%%%%%%%%%%%%%%%%%
%%%%%%%%%%%%%%%%%%%%%%%%%%%%%%%%%%%%%%%%%%%%%%%%%%%%%%
The propagation of the right- and left-moving  pump intensities along the fiber follows the following differential equation:
%%%%%%%
\begin{align}
\label{eq:pumpprop1}
\pm\dfrac{dI^\pm_p}{dz}
&=\Big[(\sigma^a_p+\sigma^e_p)N_2-\sigma^a_p N_t-\alpha_b\Big]I^\pm_p\\
\nonumber
&=-\dfrac{\sigma^a_p\,\left(1+\sum\limits^n_{j=1}\gamma_j\,\itpm_j\right)}{1+\itpm_p+\sum\limits^n_{j=1}\itpm_j}
N_t\,I^\pm_p-\alpha_b\,I^\pm_p.
\end{align}
%%%%%%%
$\alpha_b$ characterizes the undesirable pump attenuation due to scattering ($\alpha_{bs}$) and parasitic absorption ($\alpha_{ba}$),
where $\alpha_b\,=\,\alpha_{ba}+\alpha_{bs}$. 
At any point along the fiber, the total pump (signal) power is obtained by integrating the
local pump (signal) intensity over the transverse cross section. This can be formally expressed as
%%%%%%%
\begin{align}
P^\pm_j(z)\,=\,\int^{\infty}_0 (2\pi\,r\,dr)\,I^\pm_j(r,z),\quad
j=1,\cdots,n, \ {\rm or}\ j=p,
\end{align}
%%%%%%%
where we have implicitly assumed azimuthal symmetry in the fiber geometry and beam profiles. 

Earlier we stated that the pump power is uniformly distributed over the inner cladding of 
radius $b$ and the signal has a Gaussian profile. Therefore, the pump and signal intensity 
profiles can be {\em approximately} related to the powers by
%%%%%%%
\begin{align}
I^\pm_p(r,z)&=\dfrac{1}{\pi\,b^2}\,\Theta_b(r)\,P^\pm_p(z),\\
\nonumber
I^\pm_j(r,z)&=f_w(r)\,P^\pm_j(z),\quad j=1,\cdots,n.
\end{align}
%%%%%%%
Here, $f_w(r)=2\,g_w(r)/(\pi\,w^2)$ and $\int^\infty_0 (2\pi\,r\,dr)\,f_w=1$. Therefore, we can
conveniently express the pump propagation equation \ref{eq:pumpprop1} as
%%%%%%%
\begin{align}
\label{eq:pumpprop2}
\pm\dfrac{dI^\pm_p}{dz}
=-\dfrac{\Theta_a\Theta_b}{\pi\,b^2}\dfrac{\sigma^a_p\,\left(1+g_w\,\sum\limits^n_{j=1}\gamma_j\,\ptpm_j\right)}{1+\Theta_b\,\ptpm_p+g_w\,\sum\limits^n_{j=1}\ptpm_j}
N_0\,P^\pm_p-\alpha_b\,I^\pm_p,
\end{align}
%%%%%%%
where we have used the following definitions:
%%%%%%%
\begin{align}
\nonumber
&p^\pm_p(z)=P^\pm_p(z)/P^{\rm sat}_p,\ \ &&p^\pm_j(z)=P^\pm_j(z)/P^{\rm sat}_j,\ \ j=1,\cdots,n,\\
\nonumber
&P^{\rm sat}_p=I^{\rm sat}_p\,(\pi\,b^2),\ \ &&P^{\rm sat}_j=I^{\rm sat}_j\,(\pi w^2/2),\\
&i^\pm_p(r,z)=\Theta_b(r)p^\pm_p(z),\ \ &&i^\pm_j(r,z)=g_w(r)p^\pm_j(r,z).
\end{align}
%%%%%%%
If we integrate Eq.~\ref{eq:pumpprop2} over the transverse plane and use Eq.~\ref{eq:app1}, we end up with
%%%%%%%
\begin{align}
&\pm\,\dfrac{dp^\pm_p(z)}{dz}
=-\alpha_b\,p^\pm_p(z)-N_0\,\sigma^a_p\,\Gamma\,\Bigg[-\ln(1-\eta)\,\dfrac{\mathbb{B}}{\mathbb{D}}\\
\nonumber
&+\left(
\dfrac{\mathbb{A}\,\mathbb{D}-\mathbb{B}\,\mathbb{C}}{\mathbb{C}\,\mathbb{D}}
\right)
\times
\ln\left(1+\dfrac{\eta}{1-\eta}\dfrac{\mathbb{C}}{\mathbb{C}+\mathbb{D}}\right)\Bigg]\,p^\pm_p(z),
\end{align}
%%%%%%%
where 
%%%%%%%
\begin{align}
\nonumber
&\mathbb{A}=1,\quad
\mathbb{B}=\sum\limits^n_{k=1}\gamma_k\,\ptpm_k,\quad
\mathbb{C}=1+\ptpm_p,\quad 
\mathbb{D}=\sum\limits^n_{k=1}\ptpm_k,\\
&\Gamma=(\pi w^2/2)/(\pi b^2).
\label{Eq:CD}
\end{align}
%%%%%%%
%%%%%%%%%%%%%%%%%%%%%%%%%%%%%%%%%%%%%%%%%%%%%%%%%%%%%%
%%%%%%%%%%%%%%%%%%%%%%%%%%%%%%%%%%%%%%%%%%%%%%%%%%%%%%
\subsection{Signal (laser and ASE) propagation}
\label{sec:signalprop}
%%%%%%%%%%%%%%%%%%%%%%%%%%%%%%%%%%%%%%%%%%%%%%%%%%%%%%
%%%%%%%%%%%%%%%%%%%%%%%%%%%%%%%%%%%%%%%%%%%%%%%%%%%%%%
The propagation of the right- and left-moving  signal intensities along the fiber follows the following differential equation:
%%%%%%%
\begin{align}
\nonumber
\pm\dfrac{dI^\pm_j}{dz}
&=\Big[(\sigma^a_j+\sigma^e_j)N_2-\sigma^a_j N_t\Big]I^\pm_j-\widetilde{\alpha}_b\,I^\pm_j+\sigma^e_j\,N_2\,\Pi_j\,f_w\\
\nonumber
&=
\dfrac{\beta_p\,\itpm_p+\sum\limits^n_{k=1}\beta_k\,\itpm_k}{1+\itpm_p+\sum\limits^n_{k=1}\itpm_k}
N_t\Bigg((\sigma^a_j+\sigma^e_j)I^\pm_j+\sigma^e_j\,\Pi_j\,f_w\Bigg)
\\
&-N_t \sigma^a_j I^\pm_j(z) -\widetilde{\alpha_b}\,I^\pm_j,
\label{eq:signalprop1}
\end{align}
%%%%%%%
where $\Pi_j=2hc^2\delta\lambda/\lambda^3_j$.
$\widetilde{\alpha_b}$ characterizes the undesirable signal attenuation due to scattering ($\widetilde{\alpha_{bs}}$) 
and parasitic absorption ($\widetilde{\alpha_{ba}}$),
where $\widetilde{\alpha_b}\,=\,\widetilde{\alpha_{ba}}+\widetilde{\alpha_{bs}}$; the signal attenuation parameters are assumed to be the same 
for all signal wavelengths considered in this paper, because they all propagate through the fiber core.
Using the signal and pump transverse profiles, we can conveniently express the signal propagation equation \ref{eq:signalprop1} as
%%%%%%%
\begin{align}
\label{eq:signalprop2}
&\pm\dfrac{dI^\pm_j}{dz}
=-\widetilde{\alpha_b}\,I^\pm_j-N_0\, \sigma^a_j\, P^\pm_j \Theta_a\, f_w\\
\nonumber
&+
\dfrac{\Theta_b\,\beta_p\,\ptpm_p+g_w\sum\limits^n_{k=1}\beta_k\,\ptpm_k}{1+\Theta_b\ptpm_p+g_w\,\sum\limits^n_{k=1}\ptpm_k}
N_0\Bigg((\sigma^a_j+\sigma^e_j)P^\pm_j+\sigma^e_j\,\Pi_j\,\Bigg)\,
\Theta_a\,f_w,
\end{align}
%%%%%%%
where $j=1,\cdots,n$.
If we integrate Eq.~\ref{eq:signalprop2} over the transverse plane and use Eq.~\ref{eq:app2}, we end up with
%%%%%%%
\begin{align}
\nonumber
&\pm\dfrac{dp^\pm_j}{dz}
=-\widetilde{\alpha_b}\,p^\pm_j-N_0\,\eta\,\sigma^a_j\, p^\pm_j
+
N_0\Bigg((\sigma^a_j+\sigma^e_j)p^\pm_j+\sigma^e_j\,\tilde{\Pi}_j\,\Bigg)
\\
&\times
\Bigg[
\eta\,\dfrac{\widetilde{\mathbb{B}}}{\mathbb{D}}-\left(
\dfrac{\widetilde{\mathbb{A}}\,\mathbb{D}-\widetilde{\mathbb{B}}\,\mathbb{C}}{\mathbb{D}^2}
\right)\times
\ln\left(1-\eta\dfrac{\mathbb{D}}{\mathbb{C}+\mathbb{D}}\right)
\Bigg],
\end{align}
%%%%%%%
where we have 
%%%%%%%
\begin{align}
\widetilde{\mathbb{A}}=\beta_p\ptpm_p,\quad  
\widetilde{\mathbb{B}}=\sum\limits^n_{k=1}\beta_k\,\ptpm_k,\quad
\tilde{\Pi}_j=\Pi_j/P^{\rm sat}_j,
\end{align}
%%%%%%%
and $\mathbb{C}$ and $\mathbb{D}$ are defined in Eq.~\ref{Eq:CD}.
We note that in using Eq.~\ref{eq:signalprop2}, it is understood that $\Theta_b$ remains equal to unity over the domain
of integration, which is forced to be relevant only over the fiber core due to the overall $\Theta_a$ factor in the rightmost term in Eq.~\ref{eq:signalprop2}.
%%%%%%%%%%%%%%%%%%%%%%%%%%%%%%%%%%%%%%%%%%%%%%%%%%%%%%
%%%%%%%%%%%%%%%%%%%%%%%%%%%%%%%%%%%%%%%%%%%%%%%%%%%%%%
\subsection{Heat generation}
\label{sec:heat}
%%%%%%%%%%%%%%%%%%%%%%%%%%%%%%%%%%%%%%%%%%%%%%%%%%%%%%
%%%%%%%%%%%%%%%%%%%%%%%%%%%%%%%%%%%%%%%%%%%%%%%%%%%%%%
The heat density generated in the gain medium can be calculated using the net balance of energies. 
The pump and signal absorption by the ions as well as the absorption by impurities contribute to the
heating. The spontaneous emission radiates some of the power out of the fiber sides and causes
radiative cooling. The ASE also contributes to the radiative cooling by removing some of the power out
of the fiber ends. The rate of heat volume density generation $du(r,z)/dt$ at radial distance $r$ 
from the fiber center and location $z$ along the fiber is given by  
%%%%%%%
\begin{align}
\nonumber
\dfrac{du}{dt}=
&-\dfrac{dI^+_p}{dz}
+\dfrac{dI^-_p}{dz}+
\sum\limits^n_{j=1}
\left(
-\dfrac{dI^+_j}{dz}
+\dfrac{dI^-_j}{dz}
\right)-\alpha_{bs}\left(I^+_p+I^-_p\right)\\
&-\widetilde{\alpha_{bs}}\sum\limits^n_{j=1}\left(I^+_j+I^-_j\right)
-\dfrac{N_2}{\tau_{\rm r}}h\,\nu_f
+\sum\limits^n_{j=1}
2\,N_2\,\sigma^e_j\,\Pi_j\,f_w.
\end{align}
%%%%%%%

The linear heat density (per unit length) generated per unit time across the fiber cross section at location $z$ 
along the fiber is given by $Q(z)=\int^\infty_0 (2\pi\,r\,dr)\,(du(r,z)/dt)$: 
%%%%%%%
\begin{align}
\nonumber
Q(z)=Q_p+\sum\limits^n_{j=1}Q_j+Q_f,
\end{align}
%%%%%%%
where
%%%%%%%
\begin{align}
\nonumber
&Q_p=\alpha_{ba}\,\Ptpm_p(z)+N_0\,\sigma^a_p\,\Gamma\,\Bigg[-\ln(1-\eta)\,\dfrac{\mathbb{B}}{\mathbb{D}}+
\\
&\left(
\dfrac{\mathbb{A}\,\mathbb{D}-\mathbb{B}\,\mathbb{C}}{\mathbb{C}\,\mathbb{D}}
\right)
\times\ln\left(1+\dfrac{\eta}{1-\eta}\dfrac{\mathbb{C}}{\mathbb{C}+\mathbb{D}}\right)\Bigg]\,\Ptpm_p(z),
\end{align}
%%%%%%%
%%%%%%%
\begin{align}
\nonumber
&Q_j
=\widetilde{\alpha_{ba}}\,\Ptpm_j(z)+N_0\,\eta\,\sigma^a_j\,\Ptpm_j(z)
-
N_0(\sigma^a_j+\sigma^e_j)
\\
&\times
\Bigg[
\eta\,\dfrac{\widetilde{\mathbb{B}}}{\mathbb{D}}-\left(
\dfrac{\widetilde{\mathbb{A}}\,\mathbb{D}-\widetilde{\mathbb{B}}\,\mathbb{C}}{\mathbb{D}^2}
\right)\times
\ln\left(1-\eta\dfrac{\mathbb{D}}{\mathbb{C}+\mathbb{D}}\right)
\Bigg]\,\Ptpm_j(z),
\end{align}
%%%%%%%
%%%%%%%
\begin{align}
\nonumber
&Q_f=
-N_0\left(\dfrac{\pi\,w^2}{2}\right)\,\Bigg[-\ln(1-\eta)\,\dfrac{\widetilde{\mathbb{B}}}{\mathbb{D}}\\
&+\left(
\dfrac{\widetilde{\mathbb{A}}\,\mathbb{D}-\widetilde{\mathbb{B}}\,\mathbb{C}}{\mathbb{C}\,\mathbb{D}}
\right)
\times
\ln\left(1+\dfrac{\eta}{1-\eta}\dfrac{\mathbb{C}}{\mathbb{C}+\mathbb{D}}\right)\Bigg]\left(\dfrac{h\nu_f}{\tau_{\rm r}}\right),
\label{Eq:Qf}
\end{align}
%%%%%%%
where $\Ptpm_p=P_p^++P_p^-$ and $\Ptpm_j=P_j^++P_j^-$. The first term in $Q_p$ is the heat generated over the entire 
inner-cladding of the fiber; while the rest of the terms in $Q_p$ relate to the heat generated exclusively in the core, 
as is the case with $Q_j$ and $Q_f$. $Q_p$ and $Q_j$ contain parasitic absorption terms proportional to $\alpha_{ba}$
and $\widetilde{\alpha_{ba}}$, respectively. The rest of the terms in $Q_p+Q_j$ correspond to the heat generated due to 
the pump-signal quantum defect. $Q_f$ is the radiative cooling due to the anti-Stokes fluorescence.

The temperature variation in the fiber is very small as shown in~\cite{Mobini:17,wang2004thermal}. 
However, the difference between the temperature of the fiber surface and ambient can be substantial and is given by
%%%%%%%
\begin{align}
\Delta T=\dfrac{Q}{2 \pi H\,b_{o}},
\label{Eq:deltaT}
\end{align}
%%%%%%%
where $b_o$ is the radius of the whole fiber, i.e. the outer cladding and $H$ is the convective heat transfer coefficient for 
the medium surrounding the fiber. The value of $H$ is around 30\,${\rm W/m^2K}$ for air and 
1000\,${\rm W/m^2K}$ for water. The exact value of $H$ depends on the flow properties of the convection, but
these values are reasonable for the conditions in which typical fiber amplifier experiments are carried out. 
Assuming a maximum acceptable temperature rise of $\Delta T$=100\,K and $b_o$\,=\,300\,\textmu m (see next section),
the maximum acceptable value for the linear heat density is $Q_{\rm air}$\,=\,5.7\,W/m for air-cooled and 
$Q_{\rm wat}$\,=\,188\,W/m for water-cooled designs. Similarly, the maximum acceptable value for the linear heat density 
is $Q_{\rm air}$\,=\,1.18\,W/m for air-cooled and $Q_{\rm wat}$\,=\,39\,W/m for water-cooled designs, if a smaller outer cladding
radius of $b_o$\,=\,62.5\,\textmu m is used. These values will guide us in the next section to convert the calculated
heat density values for various scenarios to the actual temperature rise of the fiber.

In practice, all Yb-doped double-clad optical fibers are coated with a layer of polymer coating~\cite{Dawson:08}. For the 
steady-state condition; i.e., constant-wave (CW) amplifiers, Eq.~\ref{Eq:deltaT} is still applicable but $b_{o}$ must be 
replaced by the total fiber radius that includes the polymer coating. However, this is only true if the polymer is fully 
transparent and radiation trapping in the coating does not become a new source for heating. Of course, other common solutions
such as the use of potting material in KiloWatt-level amplifiers may not be compatible with efficient extraction of
anti-Stokes fluorescence, which is required for radiative cooling.
%%%%%%%%%%%%%%%%%%%%%%%%%%%%%%%%%%%%%%%%%%%%%%%%%%%%%%
%%%%%%%%%%%%%%%%%%%%%%%%%%%%%%%%%%%%%%%%%%%%%%%%%%%%%%
\section{Simulation of an amplifier}
\label{sec:simamp}
\begin{figure}[t]
\centering
\includegraphics[width=3.3 in]{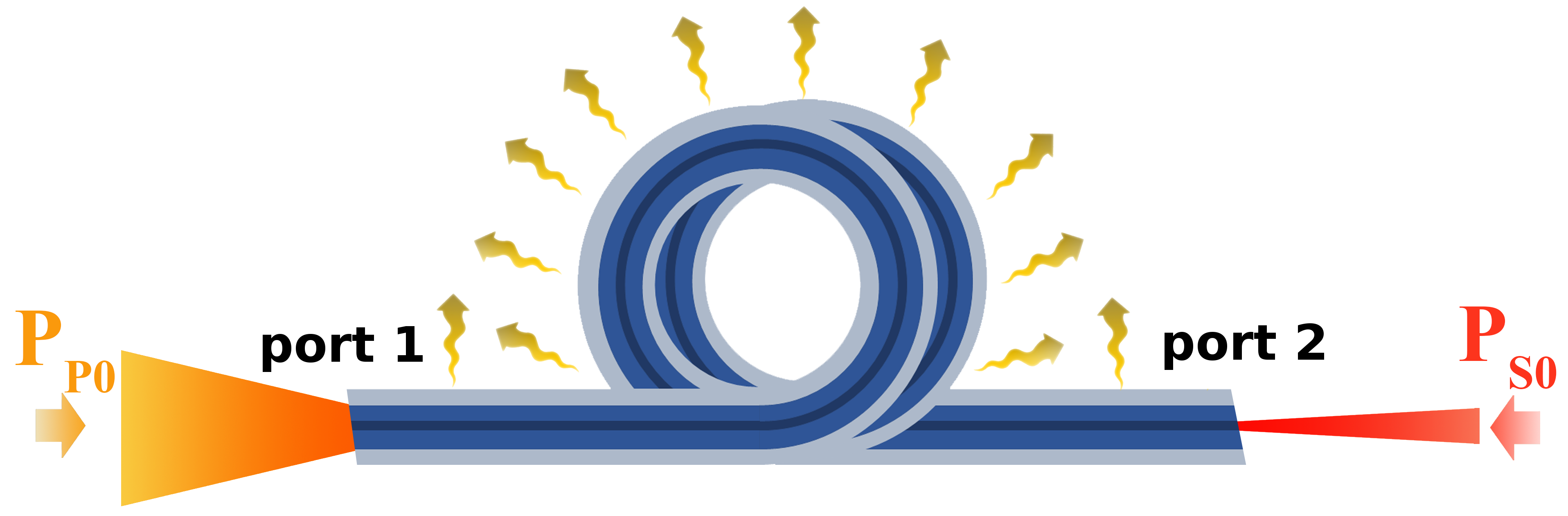}
\caption{Schematic of the fiber amplifier.}
\label{Fig:schematic}
\end{figure}

%%%%%%%%%%%%%%%%%%%%%%%%%%%%%%%%%%%%%%%%%%%%%%%%%%%%%%
%%%%%%%%%%%%%%%%%%%%%%%%%%%%%%%%%%%%%%%%%%%%%%%%%%%%%%
For the following simulation, we consider a system inspired by the amplifier described in~\cite{beier2017single}.
For the fiber geometry, we consider the core, inner cladding, and outer cladding radii
to be $a$\,=\,13\,\textmu m, $b$\,=\,240\,\textmu m, and $b_o$\,=\,300\,\textmu m, respectively.
The signal core overlap parameter is assumed to be $\eta=0.9$ and the core Yb doping density
is $N_0\,=\,1.5\times 10^{25}\,{\rm m}^{-3}$. The radiative and non-radiative lifetimes of
upper Yb manifold are taken to be $\tau_{\rm r}$\,=\,1\,ms and $\tau_{\rm nr}\,=\,10^8$\,s~\cite{paschotta1997ytterbium,digonnet2001rare}; therefore,
$\eta_q\,\approx\,1$ per Eq.~\ref{Eq:etaq}. For the absorption parameters, we assume $\alpha_{ba}$, $\alpha_{bs}$
$\widetilde{\alpha_{ba}}$, and $\widetilde{\alpha_{bs}}$ are each 5\,dB/km.
In all simulations, unless explicitly mentioned (in Fig.~\ref{Fig:heat-versus-position-1030nm-20Wpump-50mlength}), 
the amplified signal is seeded at $P_{s0}$\,=\,10\,W input power at the wavelength
of $\lambda_s$\,=\,1067\,nm (port 2, input at $z=L$). The pump is coupled to the fiber (port 1, input at $z=0$) 
in the counter-propagating direction with respect to the seed signal, where the input pump power is identified as $P_{p0}$
as it is shown in Fig.~\ref{Fig:schematic}. The fiber length is
identified as $L$ and is 35\,m long unless explicitly mentioned (in Fig.~\ref{Fig:heat-versus-position-1030nm-20Wpump-50mlength}), 
and $P_{p0}$ takes different values for different simulations. All simulations are carried out with $n$=101 signals symmetrically spaced around
$\lambda_s$\,=\,1067\,nm (corresponding to $j$\,=\,51), where each ASE component is separated by $\delta\lambda$\,=\,2\,nm.
%%%%%%
%%%%%%
%%%%%%%%%%%%%%%%%%%%%%%%%%%%%%%%%%%%%%%%%%%%%%%%%%%%%
%%%%%%%%%%%%%%%%%%%%%%%%%%%%%%%%%%%%%%%%%%%%%%%%%%%%%%
\subsection{Heat generation in a high-gain high-power amplifier}
\begin{figure}[t]
\centering
\includegraphics[width=3.3 in]{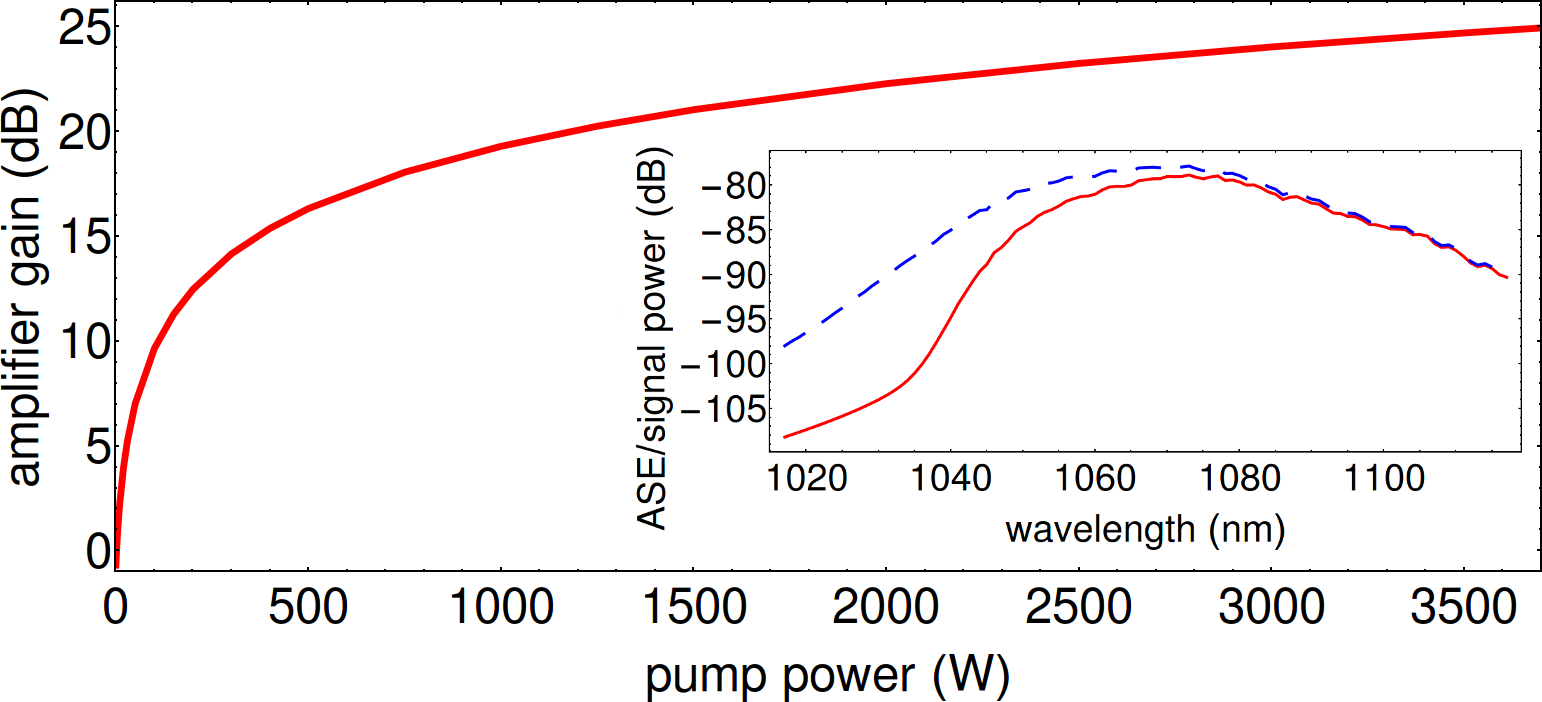}
\caption{Amplifier signal gain in dB is plotted as a function of the input pump power. 
The inset shows the ASE power spectrum relative to the amplified signal power, obtained for $P_{p0}$\,=\,3500\,W,
in dB; the solid red line is the ASE measured at port 1 and the dashed blue line
is the ASE measured at port 2. The ASE power in each spectral bin relative to the 
amplified signal.}
\label{Fig:gain-1}
\end{figure}

%%%%%%%%%%%%%%%%%%%%%%%%%%%%%%%%%%%%%%%%%%%%%%%%%%%%%%
%%%%%%%%%%%%%%%%%%%%%%%%%%%%%%%%%%%%%%%%%%%%%%%%%%%%%%
We start our discussion by considering an amplifier of length $L$\,=\,35\,m and plot the amplified signal gain
as a function of the input pump power $P_{p0}$ in Fig.~\ref{Fig:gain-1}, where $\lambda_p$\,=\,976\,nm is assumed. 
For example, for $P_{p0}$\,=\,3500\,W, the 10\,W input signal at 1067\,nm is amplified to 2950\,W, which is equivalent to 24.7\,dB of amplification.
The total pump absorption is nearly 14\,dB in this amplifier setup.
The inset in Fig.~\ref{Fig:gain-1} shows the ASE power spectrum relative to the amplified signal power for $P_{p0}$\,=\,3500\,W
in dB; the solid red line is the ASE at port 1 (ASE that copropagates with the amplified signal) and the dashed blue line
is the ASE at port 2 (ASE that copropagates with the pump). The plots show the ASE power in each spectral bin relative to the 
amplified signal in dB units; therefore, the total ASE power is sum of all the partial ASE power values in each bin.  
%%%%%%
%%%%%%

\begin{figure}[htpb]
\centering
\includegraphics[width=3.3 in]{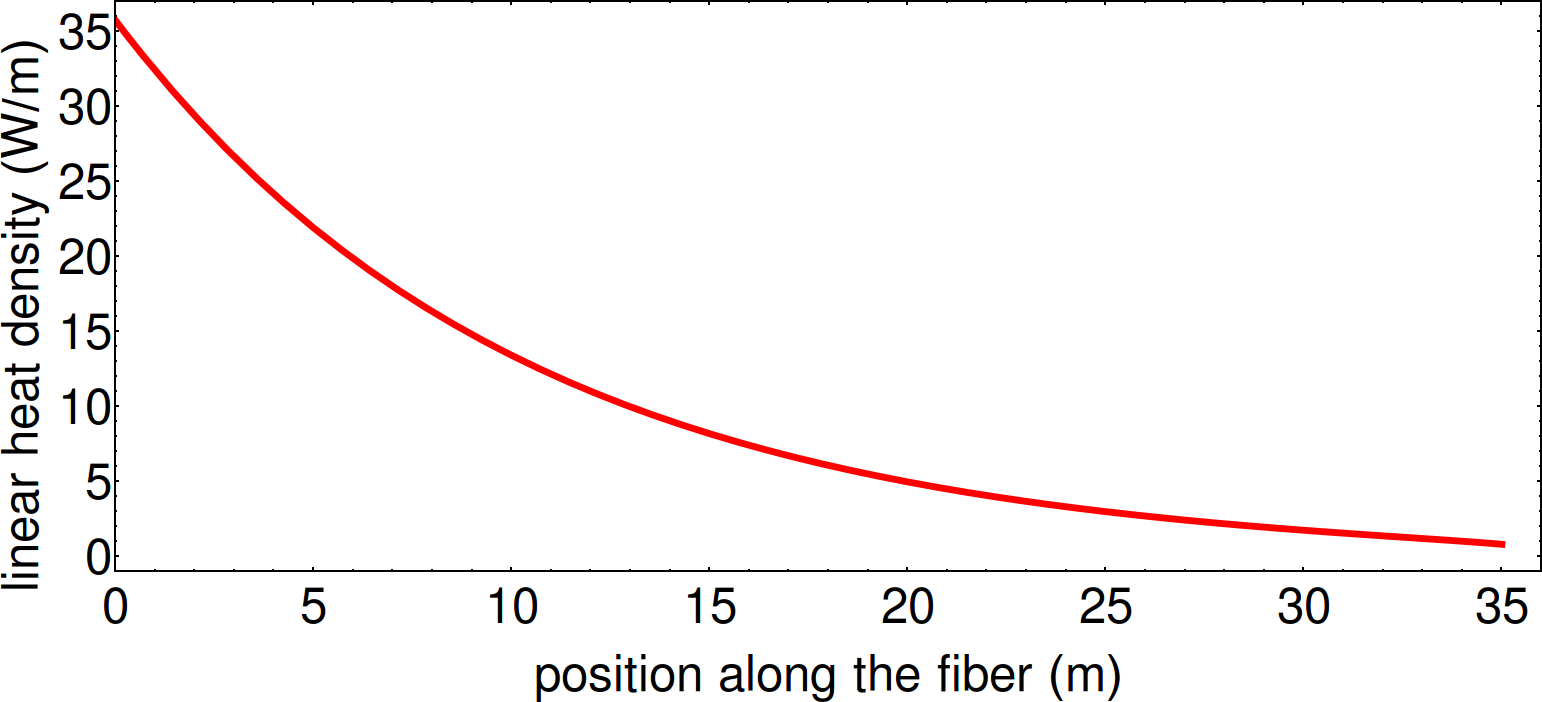}
\caption{The total linear heat density generated at each point along the fiber, plotted
for the example above at $P_{p0}$\,=\,3500\,W. $z=0$ is port 1 where the pump is injected (where the signal also takes its maximum value).}
\label{Fig:heat-3500-976-35}
\end{figure}

Figure~\ref{Fig:heat-3500-976-35} shows the total linear heat density generated at each point along the fiber, plotted
for the example above at $P_{p0}$\,=\,3500\,W. The hottest point is at port 1 where the pump is injected (where the signal also takes its maximum value). 
The large value of the liner heat density clearly shows that the fiber must be water-cooled as is also the case in~\cite{beier2017single}, 
where the fiber temperature rises $\Delta T\,\approx$\,19\,K above ambient, which is acceptable.
%%%%%%
%%%%%%

%%%%%%
\begin{figure}[t]
\centering
\includegraphics[width=3.3 in]{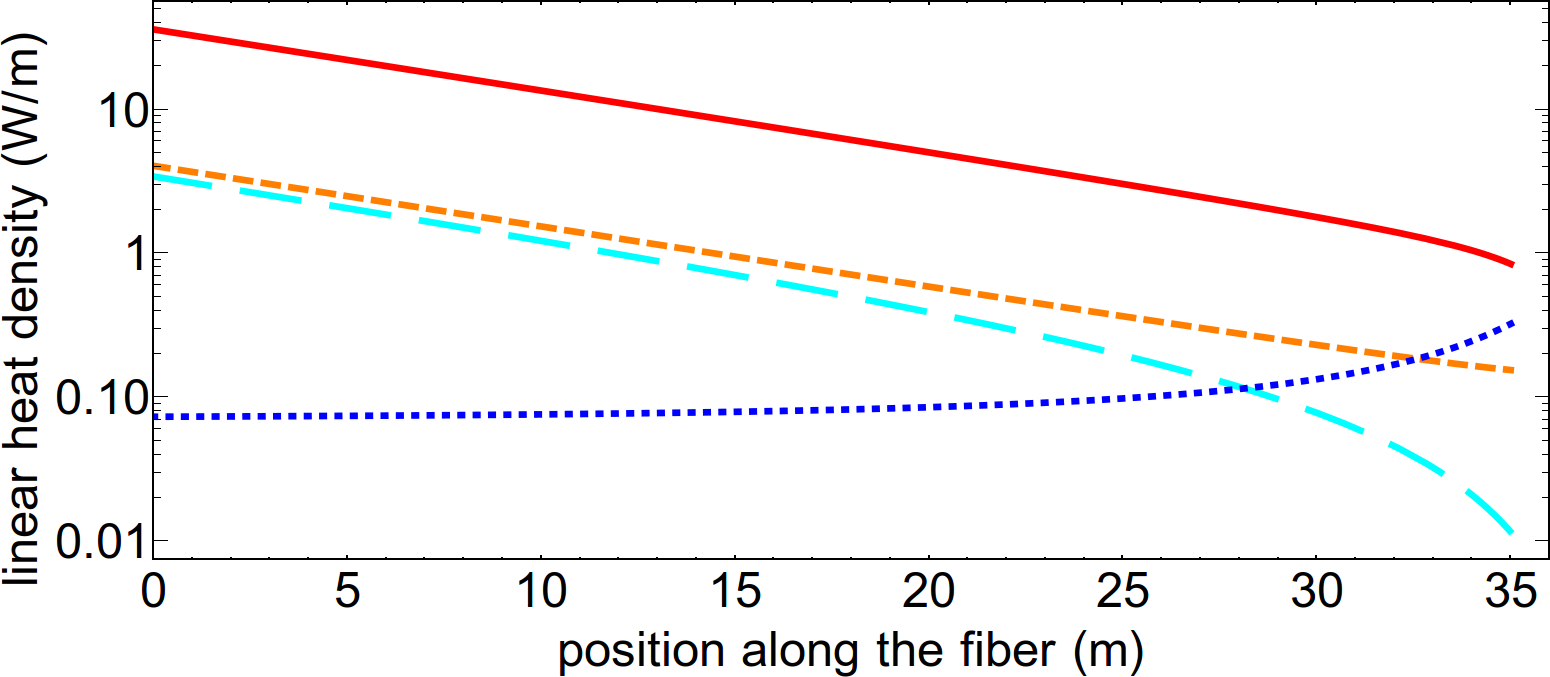}
\caption{Similar to Fig.~\ref{Fig:heat-3500-976-35} except plotted in a logarithmic scale. The solid (red) line shows 
the total heat density as in Fig.~\ref{Fig:heat-3500-976-35}, the dashed (orange) line shows the generated heat due to the pump 
parasitic absorption (proportional to $\alpha_{ba}$), the long-dashed (cyan) line shows the generated heat due to 
the laser parasitic absorption (proportional to $\widetilde{\alpha_{ba}}$), and the dotted (blue) line is the amount of radiative
cooling ($-Q_f$).}
\label{Fig:heat-3500-976-35-2}
\end{figure}
%%%%%%
In Fig.~\ref{Fig:heat-3500-976-35-2}, we show the linear heat density generated at each point along the fiber, similar to 
Fig.~\ref{Fig:heat-3500-976-35}, except plotted in a logarithmic scale. The solid (red) line shows 
the total heat density as in Fig.~\ref{Fig:heat-3500-976-35}, the dashed (orange) line shows the generated heat due to the pump 
parasitic absorption (proportional to $\alpha_{ba}$), the long-dashed (cyan) line shows the generated heat due to 
the amplified signal parasitic absorption (proportional to $\widetilde{\alpha_{ba}}$), and the dotted (blue) line is the amount of radiative
cooling ($-Q_f$) due to anti-Stokes fluorescence (note the negative sign). 
It is clear that the total generated heat in this case is dominated by the pump-signal quantum defect.
In fact, the heat generated due to the quantum defect is nearly four times larger than the heat generated due to the 
the parasitic absorptions and more than two orders of magnitude larger than the radiative cooling. In the next subsections, 
we will elaborate more on the heat generation due to quantum defect and parasitic absorption and discuss strategies to
make the radiative cooling contribution a substantial part of the heat balance. 
%%%%%%%%%%%%%%%%%%%%%%%%%%%%%%%%%%%%%%%%%%%%%%%%%%%%%%
%%%%%%%%%%%%%%%%%%%%%%%%%%%%%%%%%%%%%%%%%%%%%%%%%%%%%%
\subsection{Reducing the heat generation due to quantum defect}
\begin{figure}[t]
\centering
\includegraphics[width=3.3 in]{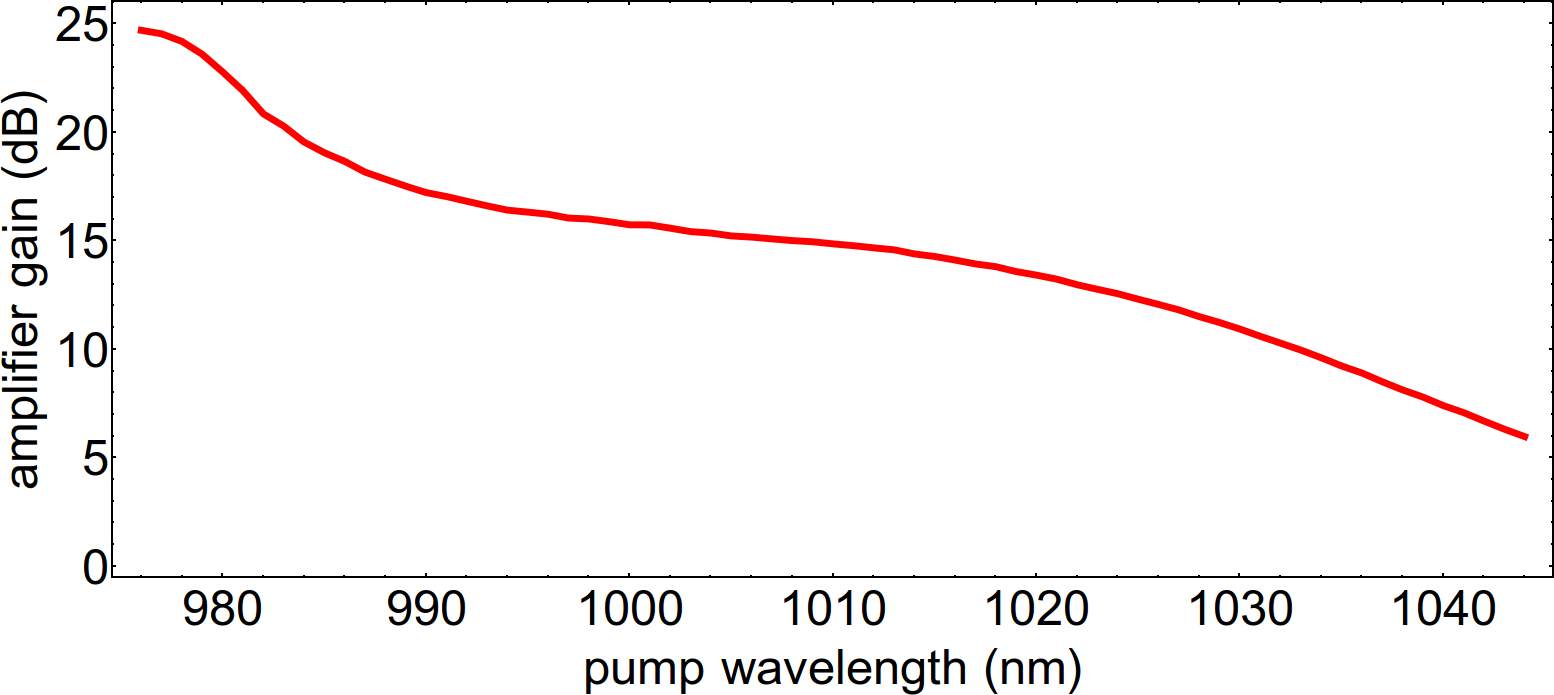}
\caption{Amplifier signal gain in dB is plotted as a function of the pump wavelength.
The input pump power is $P_{p0}$\,=\,3500\,W and the amplifier length is $L$\,=\,35\,m.}
\label{Fig:gain-vs-wavelength-3500-35}
\end{figure}

%%%%%%%%%%%%%%%%%%%%%%%%%%%%%%%%%%%%%%%%%%%%%%%%%%%%%%
%%%%%%%%%%%%%%%%%%%%%%%%%%%%%%%%%%%%%%%%%%%%%%%%%%%%%%
As we observed in the previous subsection, the heat generated due to the quantum defect totally overshadows that of the 
parasitic absorption and radiative cooling. In order to reduce the impact of the quantum defect, the pump wavelength must 
be chosen closer to the wavelength of the amplified signal. However,  $\lambda_p$\,=\,976\,nm corresponds to the peak
absorption wavelength of the pump; therefore, the amplifier gain will be reduced if the all other factors including the total
input pump power and amplifier length are unchanged. In Fig.~\ref{Fig:gain-vs-wavelength-3500-35}, we plot the amplifier signal 
gain in dB as a function of the pump wavelength. The input pump power is $P_{p0}$\,=\,3500\,W and the amplifier length 
is $L$\,=\,35\,m, and all other fiber parameters are the same as before. At each pump wavelength, the gain can be maximized 
by changing the fiber length, but the trend will be maintained as the lower efficiency is inevitable due to a lower pump 
absorption cross section at longer pump wavelength.
%%%%%%
%%%%%%

\begin{figure}[th]
\centering
\includegraphics[width=3.3 in]{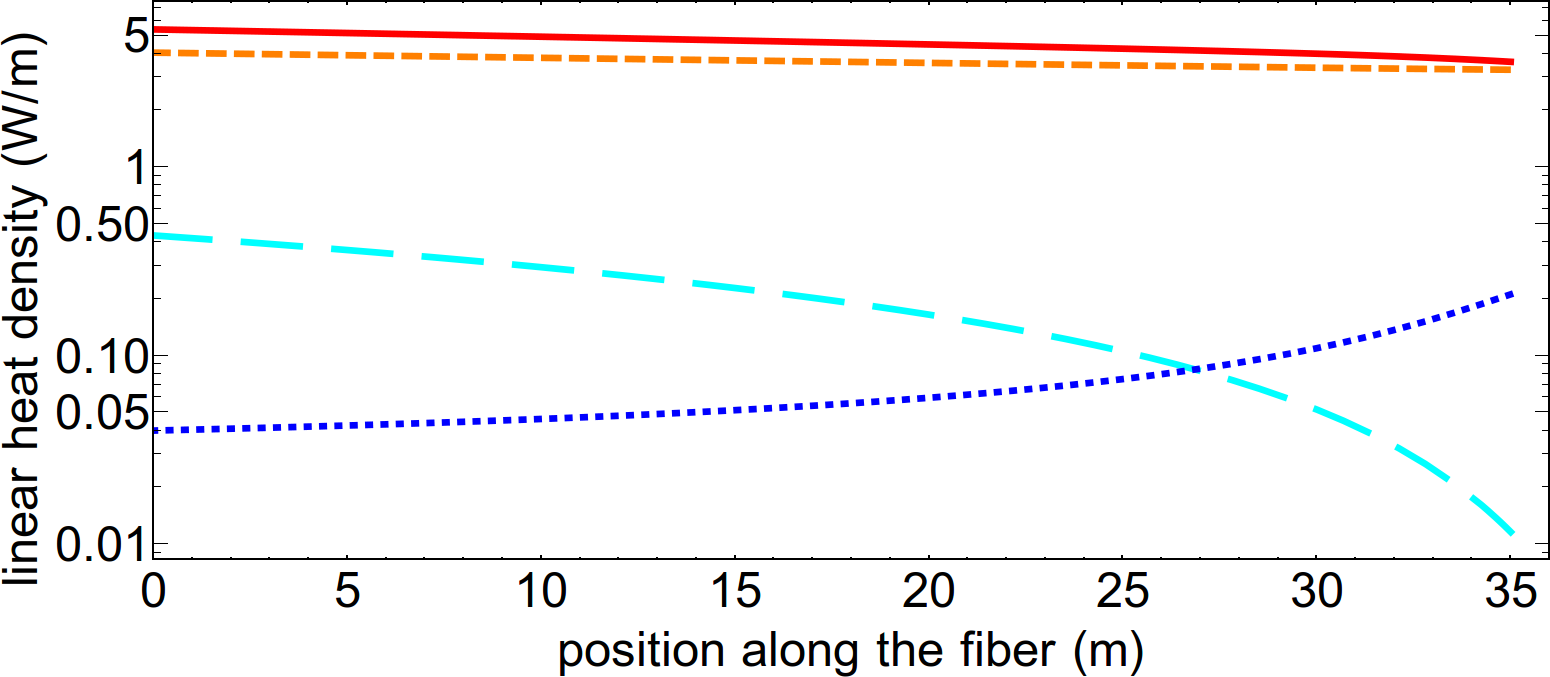}
\caption{The total linear heat density generated at each point along the fiber, plotted
for $P_{p0}$\,=\,3500\,W and $\lambda_p$\,=\,1000\,nm. $z=0$ is port 1 where the pump is injected .
The solid (red) line shows 
the total heat density as in Fig.~\ref{Fig:heat-3500-976-35}, the dashed (orange) line shows the generated heat due to the pump 
parasitic absorption (proportional to $\alpha_{ba}$), the long-dashed (cyan) line shows the generated heat due to 
the laser parasitic absorption (proportional to $\widetilde{\alpha_{ba}}$), and the dotted (blue) line is the amount of radiative
cooling ($-Q_f$).}
\label{Fig:heat-3500-1000-35-2}
\end{figure}

\begin{figure}[th]
\includegraphics[width=3.3 in]{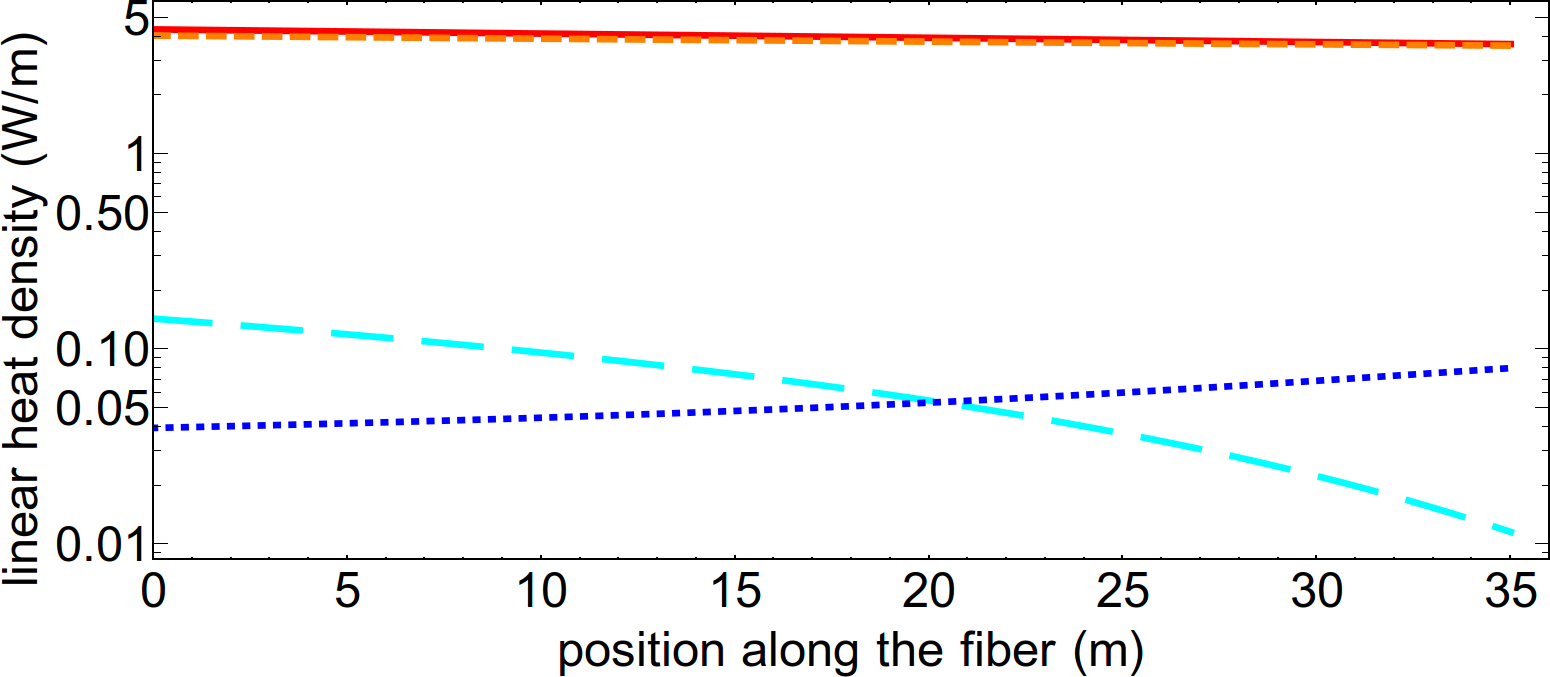}
\caption{The total linear heat density generated at each point along the fiber, plotted
for $P_{p0}$\,=\,3500\,W and $\lambda_p$\,=\,1030\,nm. $z=0$ is port 1 where the pump is injected .
The solid (red) line shows 
the total heat density as in Fig.~\ref{Fig:heat-3500-976-35}, the dashed (orange) line shows the generated heat due to the pump 
parasitic absorption (proportional to $\alpha_{ba}$), the long-dashed (cyan) line shows the generated heat due to 
the laser parasitic absorption (proportional to $\widetilde{\alpha_{ba}}$), and the dotted (blue) line is the amount of radiative
cooling ($-Q_f$).}
\label{Fig:heat-3500-1030-35-2}
\end{figure}

In order to explore the impact of the pump wavelength on the generated heat, we explore the
heat density generated along the fiber amplifier, similar to what we previously presented in 
Fig.~\ref{Fig:heat-3500-976-35-2}, except for $\lambda_p$\,=\,1000\,nm (Fig.~\ref{Fig:heat-3500-1000-35-2})
and $\lambda_p$\,=\,1030\,nm (Fig.~\ref{Fig:heat-3500-1030-35-2}). For $\lambda_p$\,=\,1000\,nm in Fig.~\ref{Fig:heat-3500-1000-35-2},
the total heat is now dominated by the parasitic absorption of the pump, which is nearly five times larger than
the heat generated due to the quantum defect. The parasitic absorption of the amplified signal is lower and is comparable in size 
to the heat generated by the quantum defect. On the other hand, the radiative cooling power is still two orders of magnitude smaller than the 
total generated heat.

%%%%%%
%%%%%%
%%%%%%
%%%%%%
For $\lambda_p$\,=\,1030\,nm in Fig.~\ref{Fig:heat-3500-1030-35-2}, the total heat is dominated by the parasitic absorption of 
the pump, similar to the case of $\lambda_p$\,=\,1000\,nm in Fig.~\ref{Fig:heat-3500-1000-35-2}. In this case, the parasitic 
absorption of the amplified signal is nearly 30 times lower than that of the pump. Similar to Fig.~\ref{Fig:heat-3500-1000-35-2},
the radiative cooling power is two orders of magnitude lower than the total generated heat. Therefore, as we expect, 
bringing the pump wavelength closer to that of the amplified signal results in a substantial reduction in the total heat 
generation due to the pump-signal quantum defect. However, other strategies must be pursued to increase the relative 
contribution of the radiative cooling due to anti-Stokes florescence.
%%%%%%%%%%%%%%%%%%%%%%%%%%%%%%%%%%%%%%%%%%%%%%%%%%%%%%
%%%%%%%%%%%%%%%%%%%%%%%%%%%%%%%%%%%%%%%%%%%%%%%%%%%%%%
\subsubsection{Impact of the total pump absorption versus wavelength}
%%%%%%%%%%%%%%%%%%%%%%%%%%%%%%%%%%%%%%%%%%%%%%%%%%%%%%
%%%%%%%%%%%%%%%%%%%%%%%%%%%%%%%%%%%%%%%%%%%%%%%%%%%%%%
%%%%%%
\begin{figure}[hbt]
\centering
\includegraphics[width=3.3 in]{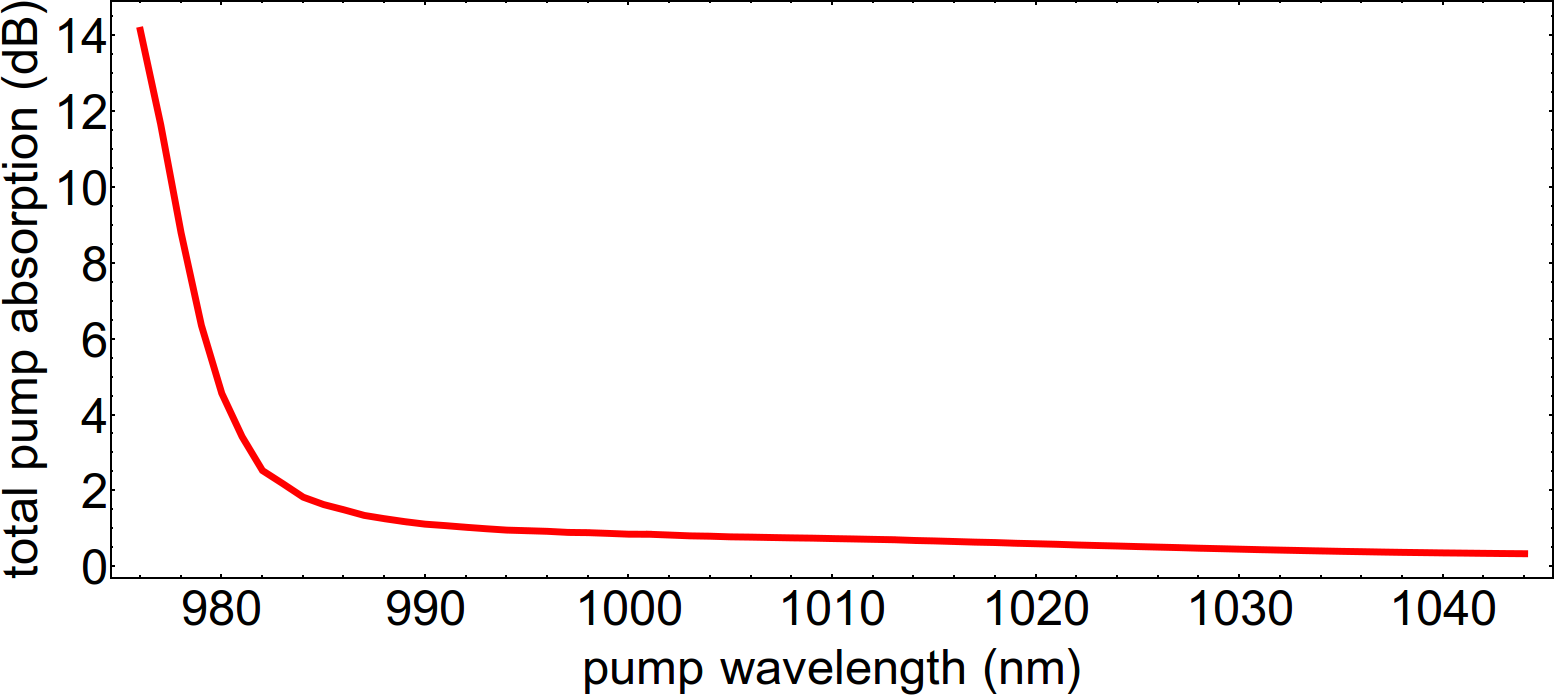}
\caption{Pump absorption in dB is plotted as a function of the pump wavelength. 
The input pump power is $P_{p0}$\,=\,3500\,W and the amplifier length is $L$\,=\,35\,m.
Note that for the reported pump absorption, the contribution of the undesirable pump attenuation, which is proportional 
to $\alpha_b$, is removed.}
\label{Fig:pump-absorption-1}
\end{figure}
%%%%%%
An important reason behind the rapid drop in the signal gain for larger wavelengths in Fig.~\ref{Fig:gain-vs-wavelength-3500-35} 
is the small pump absorption due to the decreasing absorption cross section in larger wavelengths. In Fig.~\ref{Fig:pump-absorption-1},
we plot the total pump absorption in dB as a function of the pump wavelength. 
Similar to Fig.~\ref{Fig:gain-vs-wavelength-3500-35}, the input pump power is $P_{p0}$\,=\,3500\,W and the amplifier length is $L$\,=\,35\,m.
Note that for the reported pump absorption, the contribution of the undesirable pump attenuation, which is proportional 
to $\alpha_b$, is removed. One method to address this problem is to increase the length of the fiber, such that more pump is absorbed along the fiber.

%%%%%%
\begin{figure}[hbt]
\centering
\includegraphics[width=3.3 in]{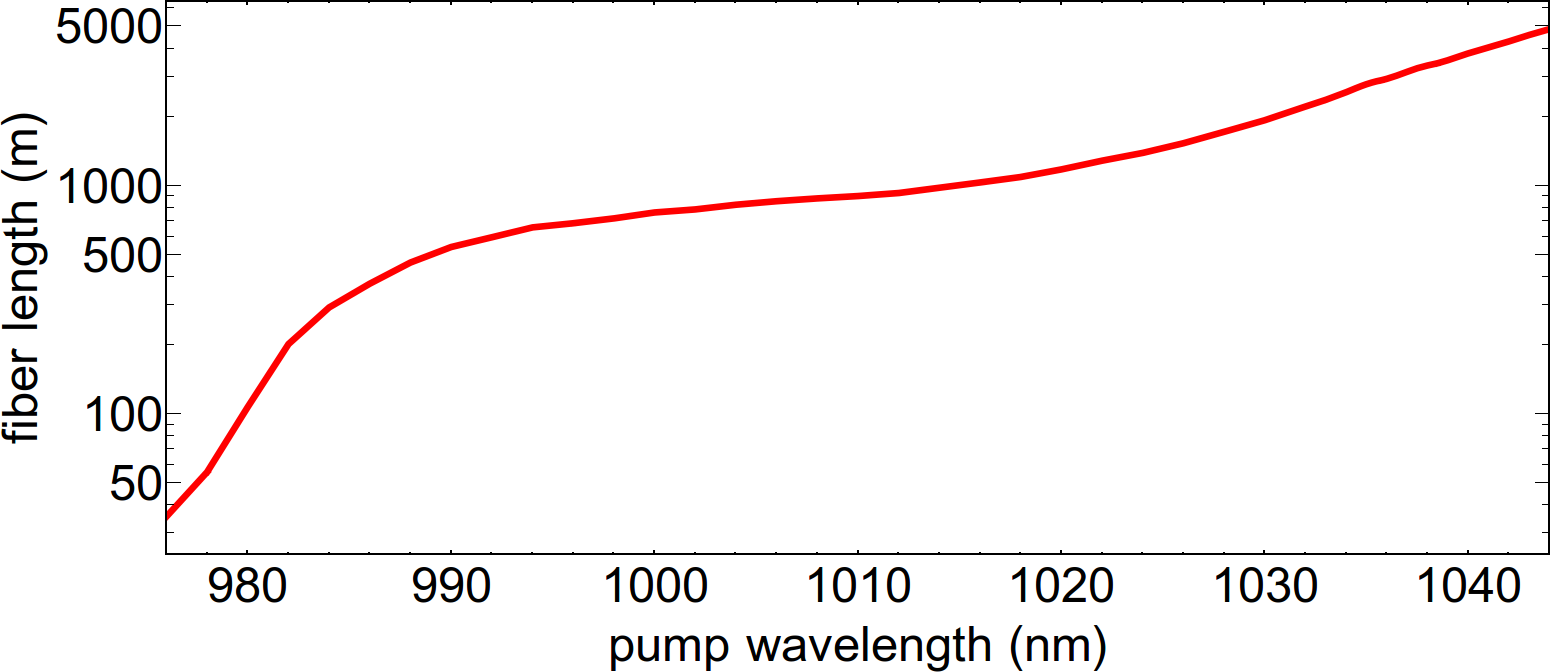}
\caption{The required fiber length $L$ to obtain 14\,dB of pump absorption is plotted, excluding the
pump attenuation due to $\alpha_b$, as a function of the pump wavelength.}
\label{Fig:length-for-14-dB-pump-attenuation}
\end{figure}
%%%%%%
%%%%%%
\begin{figure}[htb]
\centering
\includegraphics[width=3.3 in]{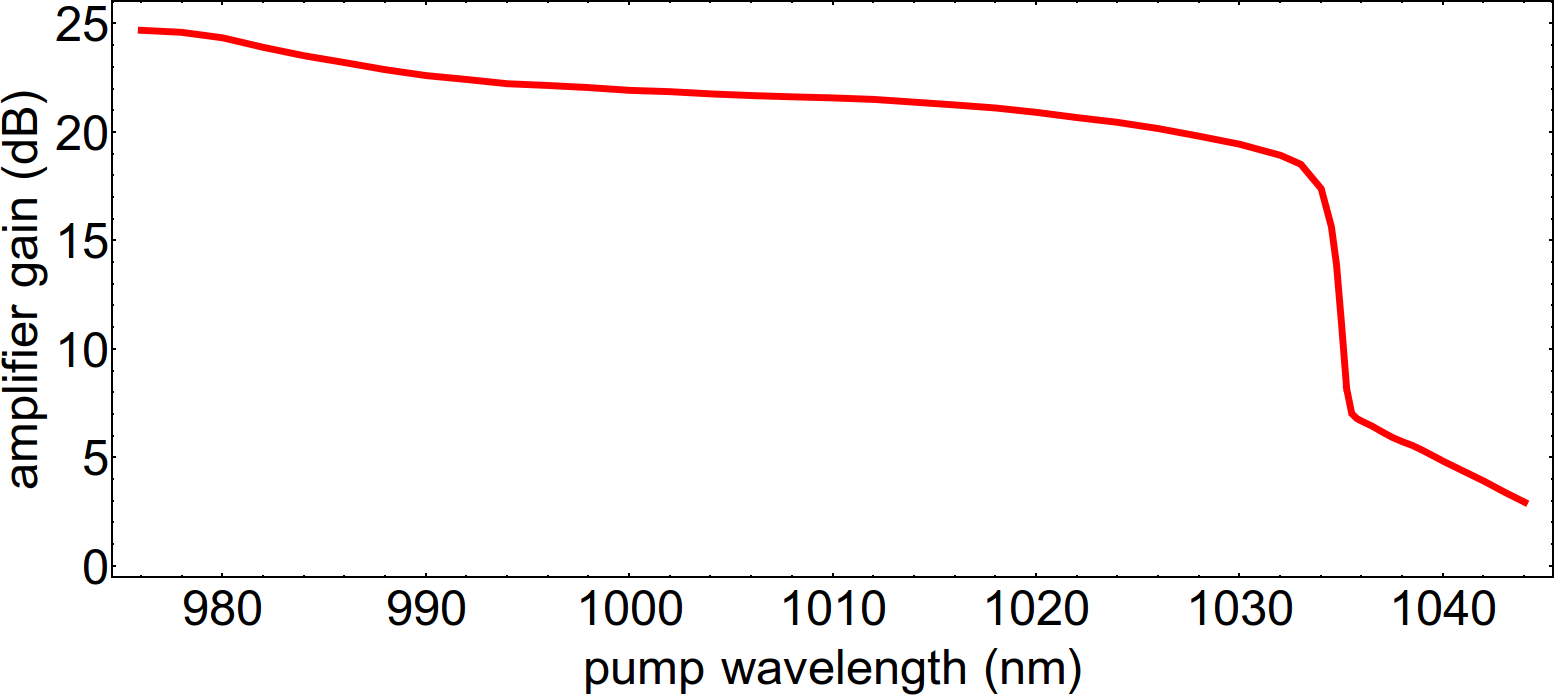}
\caption{Amplifier signal gain in dB is plotted as a function of the pump wavelength.
The input pump power is $P_{p0}$\,=\,3500\,W and the amplifier length is chosen according 
to Fig.~\ref{Fig:length-for-14-dB-pump-attenuation}, such that the pump absorption is 14\,dB, excluding the
pump attenuation due to $\alpha_b$.}
\label{Fig:signal-gain-for-14-dB-pump-attenuation}
\end{figure}
%%%%%%
In Fig.~\ref{Fig:length-for-14-dB-pump-attenuation}, we show the required fiber length to obtain 14\,dB of pump absorption, excluding the
pump attenuation due to $\alpha_b$, as a function of the pump wavelength. Of course, in practice one needs to worry about undesirable nonlinear effects
that arise from the longer fiber lengths~\cite{Dajani:08,BallatoDragic}. 
In Fig.~\ref{Fig:signal-gain-for-14-dB-pump-attenuation}, we plot the amplifier signal gain in dB as a function of 
the pump wavelength, where the fiber length is chosen according to Fig.~\ref{Fig:length-for-14-dB-pump-attenuation} for 14\,dB of pump absorption. 
Again, the input pump power is $P_{p0}$\,=\,3500\,W and the amplified signal is seeded at $P_{s0}$\,=\,10\,W input power at the wavelength
of $\lambda_s$\,=\,1067\,nm. This figure should be compared with Fig.~\ref{Fig:gain-vs-wavelength-3500-35} for which the amplifier length was fixed at 
$L\,=\,$35\,m. It is clear that the amplifier signal gain in Fig.~\ref{Fig:signal-gain-for-14-dB-pump-attenuation} is considerably higher than 
Fig.~\ref{Fig:gain-vs-wavelength-3500-35} for most values of wavelength. However, for $\lambda_p\gtrsim 1034$\,nm, the amplifier signal gain 
rapidly drops in Fig.~\ref{Fig:signal-gain-for-14-dB-pump-attenuation}. The reason behind this rapid drop is the extremely low pump absorption 
cross section necessitates an exponential growth in the required amplifier to maintain 14\,dB of pump absorption 
in Fig.~\ref{Fig:length-for-14-dB-pump-attenuation} (note the logarithmic vertical scale); therefore, the undesirable signal attenuation 
proportional to $\widetilde{\alpha_{b}}$ rapidly takes over for $\lambda_p\gtrapprox 1034$\,nm and the signal gain drops accordingly.

%%%%%%
\begin{figure}[htb]
\centering
\includegraphics[width=3.3 in]{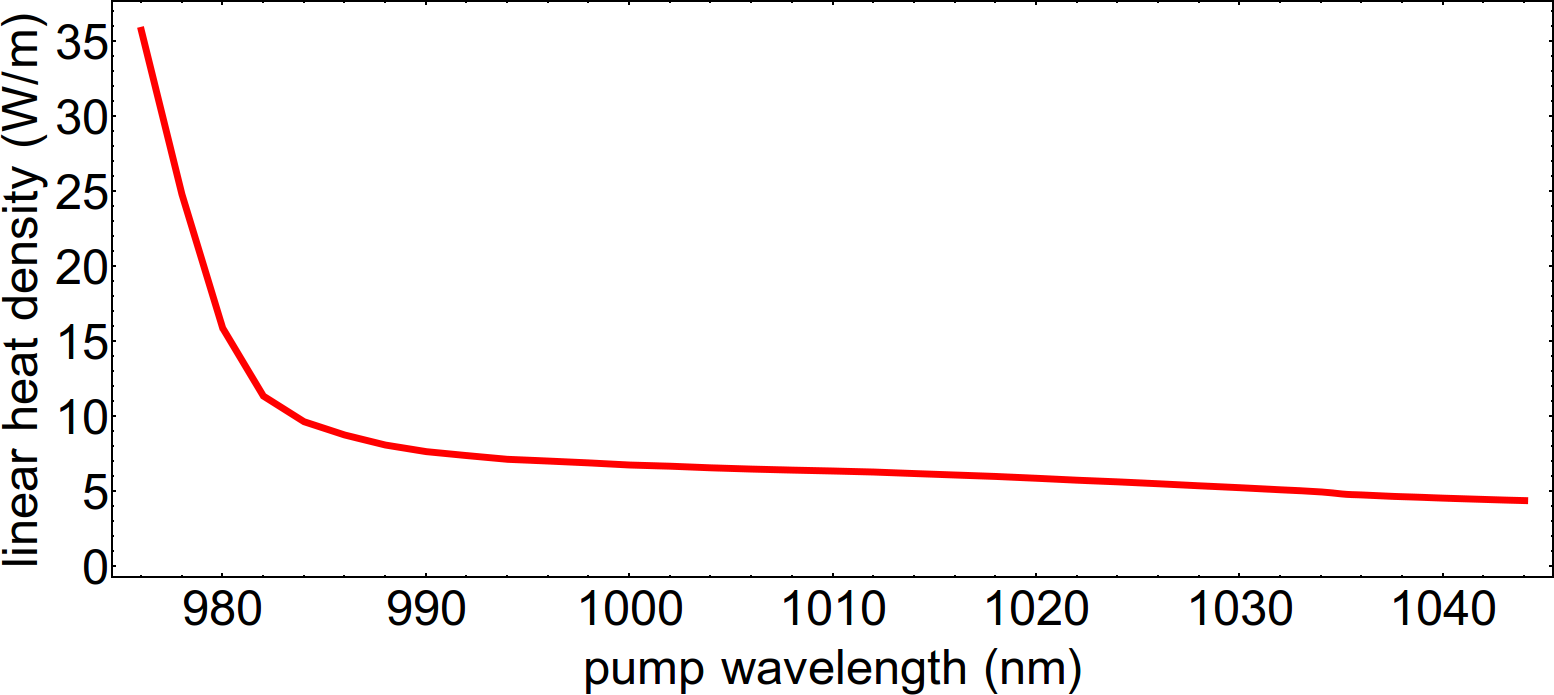}
\caption{Linear heat density generated in port 1 ($z\,=\,0$, which is the hottest point in the fiber) 
is plotted as a function of the pump wavelength. The input pump power is $P_{p0}$\,=\,3500\,W and the amplifier length is chosen according 
to Fig.~\ref{Fig:length-for-14-dB-pump-attenuation}. The amplifier signal gain for this plot is shown in Fig.~\ref{Fig:signal-gain-for-14-dB-pump-attenuation}.}
\label{Fig:heat-for-14-dB-pump-attenuation}
\end{figure}
%%%%%%
In Fig.~\ref{Fig:heat-for-14-dB-pump-attenuation}, we plot the linear heat density in port 1 ($z\,=\,0$, which is the hottest point in the fiber) 
as a function of the pump wavelength for the amplifier design related to Figs.~\ref{Fig:length-for-14-dB-pump-attenuation} 
and~\ref{Fig:signal-gain-for-14-dB-pump-attenuation}. It can be seen that as the pump wavelength increases, the pump absorption cross section decreases; 
therefore the local value of the linear heat density decreases. The results presented in these figures show 
that $1020\,{\rm nm}\lesssim \lambda_p\lesssim  1030\,{\rm nm}$ may be considered as the desirable pump wavelength range in high-power amplifiers
if heat mitigation is an issue, because the generated linear heat density is relatively low, radiative cooling can potentially 
contribute to heat reduction, while the signal gain can still be appreciable for a sufficiently long fiber. However, undesirable nonlinear 
effects must still be mitigated in such fibers~\cite{Dajani:08,BallatoDragic}.
%%%%%%%%%%%%%%%%%%%%%%%%%%%%%%%%%%%%%%%%%%%%%%%%%%%%%%
%%%%%%%%%%%%%%%%%%%%%%%%%%%%%%%%%%%%%%%%%%%%%%%%%%%%%%
\subsection{Reducing the heat generation due to parasitic absorption}
%%%%%%%%%%%%%%%%%%%%%%%%%%%%%%%%%%%%%%%%%%%%%%%%%%%%%%
%%%%%%%%%%%%%%%%%%%%%%%%%%%%%%%%%%%%%%%%%%%%%%%%%%%%%%
The parasitic absorption heat generated by the pump and signal is given by $\alpha_{ba}\,\Ptpm_p$ and 
$\widetilde{\alpha_{ba}}\,\Ptpm_j$, respectively. In order for radiative cooling to have a chance 
in reducing the temperature of the fiber amplifier, the parasitic absorption heat must be comparable 
in size to the radiative cooling term $Q_f$ (assuming that the contribution due to quantum defect is minimal). 
A useful estimate of the radiative cooling can be obtained 
from Eq.~\ref{Eq:Qf} in the limit where the signal and/or the pump intensity is much larger than the
corresponding saturation value. This limit is quite reasonable and is readily satisfied due to the small 
transverse cross sections in fiber amplifies. This estimated radiative cooling power is given by 
%%%%%%%
\begin{align}
Q_f\approx -N_0\,\pi\,a^2\,\dfrac{h\nu_f}{\tau_{\rm r}}\,\beta_p
\label{Eq:Qf2}
\end{align}
%%%%%%%
if the normalized pump intensity (relative to the pump saturation intensity) dominates over the 
normalized signal intensity. If the normalized signal intensity dominates, $\beta_p$ much be replaced by the
$\beta_s$ ($\beta_{j=51}$) corresponding to the amplified signal.  

For $\lambda_p$\,=\,1030\,nm and $\lambda_s$\,=\,1067\,nm, $Q_f$ is estimated to be $-0.106\,\rm{W/m}$ when the
normalized pump intensity is dominant and $-0.022\,\rm{W/m}$ when the normalized signal intensity is dominant.
Using the previously stated values of $\alpha_{ba}$ and $\widetilde{\alpha_{ba}}$, a maximum estimate for the
pump and the amplified signal powers is near 20\,W or 90\,W, depending on which estimate for $Q_f$ is considered (depending on the amplifier design).
These values are substantially different from the pump and signal powers of the high-power amplifier studied in the previous section.
Therefore, no optimization of the amplifier design would have resulted in any respectable radiative cooling effect, simply due to the 
large size of the pump and signal powers. In order to design an amplifier with considerable radiative cooling, either
the parasitic absorption coefficients need to be reduced, or the amplifier must operate at lower power values. The first option 
is a fiber material design issue and is outside the scope of the present paper. Therefore, we
focus on changing our designs to the case of an amplifier with lower power values to explore the radiative cooling effect.  
We note that the values of $\alpha_{ba}$ and $\widetilde{\alpha_{ba}}$ used in this paper are comparable to the values in 
conventional Yb-doped optical fibers. 

As an example, consider the same fiber discussed before, except the outer cladding radius is reduced to $b$\,=\,62.5\,\textmu m
and the fiber length is increased to $L$\,=\,50\,m. The input pump power is $P_{p0}$\,=\,20\,W, and the amplified signal is seeded 
at $P_{s0}$\,=\,0.1\,W, and is amplified to 6.97\,W at more than 18\,dB of gain. The reduced powers are intended to decrease 
the parasitic absorption powers, while the reduced cladding size is aimed at increasing the pump intensity to maintain an adequate population inversion. 

\begin{figure}[htpb]
\centering
\includegraphics[width=3.3 in]{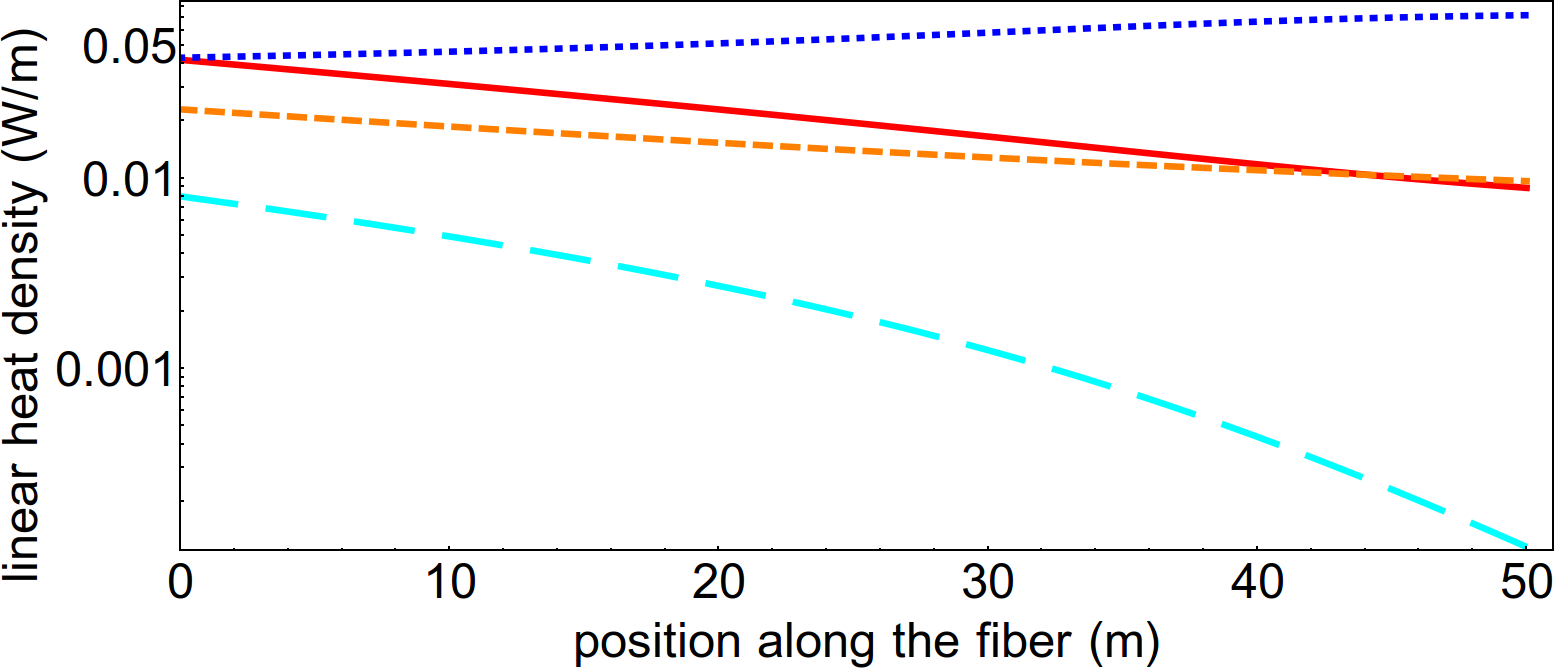}
\caption{Similar to Fig.~\ref{Fig:heat-3500-1030-35-2}, where $\lambda_p$\,=\,1030\,nm, except the cladding diameter is reduced to 
$2b$\,=\,125\,\textmu m, $P_{p0}$\,=\,20\,W, and $P_{s0}$\,=\,0.1\,W.}
\label{Fig:heat-versus-position-1030nm-20Wpump-50mlength}
\end{figure}

In Fig.~\ref{Fig:heat-versus-position-1030nm-20Wpump-50mlength}, we show the linear heat density generated at each point along the 
fiber, similar to Fig.~\ref{Fig:heat-3500-976-35-2}. It is 
observed in this case that the heat due to the pump-signal quantum defect, parasitic absorption, and radiative cooling are all 
comparable in size (up to a factor of two); therefore, radiative cooling plays an important role in reducing the heat-load 
in the fiber amplifier.
%%%%%%
%%%%%%

In order to enhance the effect of radiative cooling, one may consider to increase the core area, as well as the 
density of Yb ions $N_0$. The core area considered in this paper is nearly the maximum value that can be tolerated in conventional
fibers, before higher order modes kick in and lower the beam quality or create modal instability. Of course, customized solutions
may be applied to remove such higher order modes, but these would become highly specialized designs that are beyond the scope of this paper.
Increasing the density of Yb ions can result in increased rate for non-radiative depletion of upper energy states due to 
the self-quenching 
effect~\cite{hehlen2007model,digonnet2001rare,hoyt2003advances,barua2008influences,van1983nonradiative,auzel2003radiation,boulon2008so,nguyen2012all};
hence decreasing the quantum efficiency $\eta_q$, which will be the subject of next subsection.
%%%%%%%%%%%%%%%%%%%%%%%%%%%%%%%%%%%%%%%%%%%%%%%%%%%%%%
%%%%%%%%%%%%%%%%%%%%%%%%%%%%%%%%%%%%%%%%%%%%%%%%%%%%%%
\subsection{Impact of non-ideal quantum efficiency}
%%%%%%%%%%%%%%%%%%%%%%%%%%%%%%%%%%%%%%%%%%%%%%%%%%%%%%
%%%%%%%%%%%%%%%%%%%%%%%%%%%%%%%%%%%%%%%%%%%%%%%%%%%%%%
In practice, $\eta_q$ is never equal to unity. In order to achieve net radiative cooling~\cite{seletskiy2010laser}
or achieve perfect radiation balancing~\cite{bowman2010minimizing},
$\eta_q$ must be very close to unity. For fiber lasers and amplifiers, the excellent quantum efficiency of Yb-doped ZBLAN,
which has been measured to be as high as 0.995~\cite{DenisZBLAN,gosnell1998laser,gosnell1999laser,Peysokhan:18}, makes ZBLAN
the ideal host for radiation balancing. However, in many practical applications it may not be necessary for anti-Stokes fluorescence
to completely cancel out the heat generated from other sources. Rather, it may be sufficient for the anti-Stokes fluorescence
to only lower the heat-load to tolerable limits for the required design. In the formalism presented in this paper, $\eta_q\,<\,1$
is due to the presence of a relatively fast non-radiative decay time. 

We note that $\tau_f$ appears primarily in saturation intensities, while $\tau_{\rm r}$ appears in the expression for $Q_f$ 
(radiative cooling due to the anti-Stokes fluorescence) in Eq.~\ref{Eq:Qf}. In order to see the impact of $\eta_q\,<\,1$, consider
{\em hypothetically} a scenario where $\tau_f$ remains fixed, while lowering $\eta_q$ results in an increased value of $\tau_r$, 
which in turn lowers the impact of $Q_f$ in Eq.~\ref{Eq:Qf} proportionally. However, this is just a {\em hypothetical} example 
helping to visualize the issue more clearly. In reality, $\tau_{\rm r}$ is usually fixed and impurities lower $\tau_{\rm nr}$, which in turn
lower $\tau_f$ and result in $\eta_q\,<\,1$. However, the overall effect on the heat balance is similar in nature.
 
\begin{figure}[htpb]
\centering
\includegraphics[width=3.3 in]{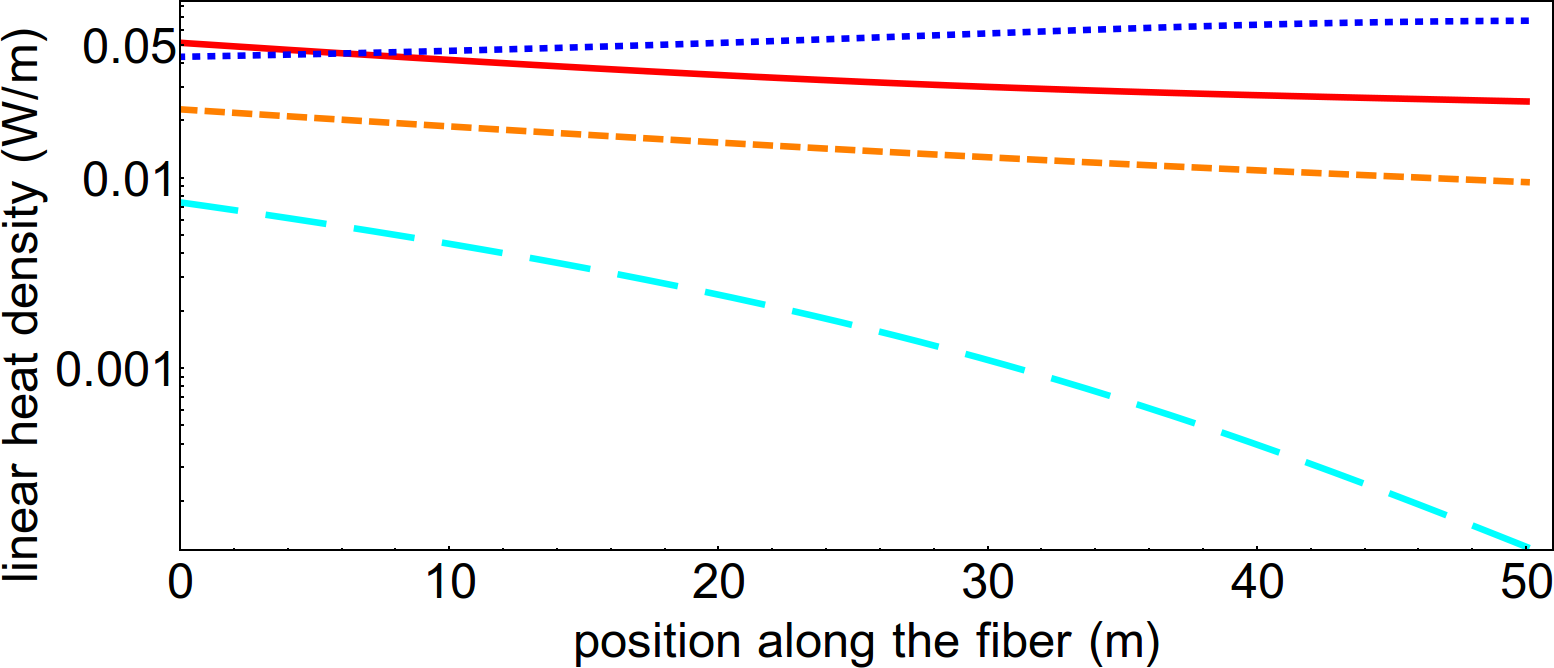}
\caption{Similar to Fig.~\ref{Fig:heat-versus-position-1030nm-20Wpump-50mlength}, except with $\eta_q\,=\,0.8$.}
\label{Fig:eta-80-heat-versus-position-1030nm-20Wpump-50mlength}
\end{figure}

In Fig.~\ref{Fig:eta-80-heat-versus-position-1030nm-20Wpump-50mlength}, we repeat the simulation of 
Fig.~\ref{Fig:heat-versus-position-1030nm-20Wpump-50mlength}, except with $\eta_q\,=\,0.8$. 
We assume that $\tau_r$\,=\,1\,ms, $\tau_{\rm nr}\,=\,4$\,ms, and $\tau_{\rm f}\,=\,0.8$\,ms, resulting
in $\eta_q\,=\,0.8$. These values may be compared with those used in previous simulations in this paper, where 
$\tau_f\,\approx\,\tau_{\rm r}$\,=\,1\,ms were used in conjunction with $\tau_{\rm nr}\,=\,10^8$\,s resulting in $\eta_q\,\approx\,1.0$.
The impact of a relatively lower $Q_f$ (compared with other heat curves) due to a smaller $\tau_{\rm f}$ (lower $\eta_q$) is clear when comparing 
Fig.~\ref{Fig:eta-80-heat-versus-position-1030nm-20Wpump-50mlength} to
Fig.~\ref{Fig:heat-versus-position-1030nm-20Wpump-50mlength}, specifically for the dotted (blue) line, which is the amount of radiative
cooling ($-Q_f$) due to anti-Stokes fluorescence. Therefore, the solid (red) line relating to the total heat density is increased relative 
to other heat curves in Fig.~\ref{Fig:eta-80-heat-versus-position-1030nm-20Wpump-50mlength} in comparison to
Fig.~\ref{Fig:heat-versus-position-1030nm-20Wpump-50mlength} (this is very clear when the solid (red) line is compared with the dashed (orange) line). 
Therefore, it is observed that while the relative impact
of the radiative cooling is weakened when $\eta_q\,<\,1$, it can still have a sizable impact on the overall heat balance. 
%%%%%%
%%%%%%
%%%%%%%%%%%%%%%%%%%%%%%%%%%%%%%%%%%%%%%%%%%%%%%%%%%%%%
%%%%%%%%%%%%%%%%%%%%%%%%%%%%%%%%%%%%%%%%%%%%%%%%%%%%%%
\section{Conclusions}
%%%%%%%%%%%%%%%%%%%%%%%%%%%%%%%%%%%%%%%%%%%%%%%%%%%%%%
%%%%%%%%%%%%%%%%%%%%%%%%%%%%%%%%%%%%%%%%%%%%%%%%%%%%%%
We presented a detailed analytical formalism intended for the thermal modeling and heat mitigation including radiative cooling for 
high-power double-clad fiber amplifiers. The formalism takes into account the spatial profile of the amplified signal and the pump 
in the double-clad geometry, the presence of the ASE, and the possibility of radiative cooling. The formalism is applied to the
analysis of a high-power Yb-doped silica fiber amplifier inspired by published experimental results. It is observed that the
pump-signal quantum defect is the dominant source of heat generation in the KiloWatt-level amplifier when the amplifier is pumped
at 976\,nm wavelength. As the pump wavelength is increased to reduce the quantum defect, the parasitic absorption of the pump 
(and amplified signal) dominate the heat generation. In this case, if the pump wavelength is longer than the mean florescence 
wavelength, radiative cooling can provide notable cooling only if the pump and signal powers are reduced to several tens of watts
or lower. This can be achieved by using a (1) smaller inner cladding in order to increase the local pump intensity to achieve an adequate
population inversion, and (2) longer fiber to increase the total signal gain. 
We also explored the impact of the non-ideal quantum efficiency of the gain material. The formalism presented here can be readily used to 
design fiber amplifiers and lasers for optimal heat mitigation, especially due to radiative cooling.  
%%%%%%%%%%%%%%%%%%%%%%%%%%%%%%%%%%%%%%%%%%%%%%%%%%%%%%
%%%%%%%%%%%%%%%%%%%%%%%%%%%%%%%%%%%%%%%%%%%%%%%%%%%%%%
\section{Appendix}
\label{sec:appendix}
%%%%%%%%%%%%%%%%%%%%%%%%%%%%%%%%%%%%%%%%%%%%%%%%%%%%%%
%%%%%%%%%%%%%%%%%%%%%%%%%%%%%%%%%%%%%%%%%%%%%%%%%%%%%%
We use the following identities in this paper:
%%%%%%%%%%%%
\begin{align}
\label{eq:app1}
&\dfrac{2}{\pi\,w^2}\int^{a}_0 (2\pi\,r\,dr)\,\dfrac{\mathbb{A}+\mathbb{B}\,g_w(r)}{\mathbb{C}+\mathbb{D}\,g_w(r)}=\\
\nonumber
&-\ln(1-\eta)\,\dfrac{\mathbb{B}}{\mathbb{D}}
+
\left(
\dfrac{\mathbb{A}\,\mathbb{D}-\mathbb{B}\,\mathbb{C}}{\mathbb{C}\,\mathbb{D}}
\right)
\times
\ln\left(1+\dfrac{\eta}{1-\eta}\dfrac{\mathbb{C}}{\mathbb{C}+\mathbb{D}}\right).
\end{align}
%%%%%%%%%%%%
%%%%%%%%%%%%
\begin{align}
\nonumber
&\int^{a}_0 (2\pi\,r\,dr)\,f_w(r)\,\dfrac{\mathbb{A}+\mathbb{B}\,g_w(r)}{\mathbb{C}+\mathbb{D}\,g_w(r)}=\\
\label{eq:app2}
&\eta\,\dfrac{\mathbb{B}}{\mathbb{D}}-\left(
\dfrac{\mathbb{A}\,\mathbb{D}-\mathbb{B}\,\mathbb{C}}{\mathbb{D}^2}
\right)\times
\ln\left(1-\eta\dfrac{\mathbb{D}}{\mathbb{C}+\mathbb{D}}\right).
\end{align}
%%%%%%%%%%%%
$\eta$ is the fractional signal power in the core and is given by
%%%%%%%%%%%%
\begin{align} 
\eta=1-\exp(-2a^2/w^2).
\end{align}
%%%%%%%%%%%%
%%%%%%%%%%%%%%%%%%%%%%%%%%%%%%%%%%%%%%%%%%%%%%%%%%%%%%
%%%%%%%%%%%%%%%%%%%%%%%%%%%%%%%%%%%%%%%%%%%%%%%%%%%%%%
\section{Funding Information}
%%%%%%%%%%%%%%%%%%%%%%%%%%%%%%%%%%%%%%%%%%%%%%%%%%%%%%
%%%%%%%%%%%%%%%%%%%%%%%%%%%%%%%%%%%%%%%%%%%%%%%%%%%%%%
This material is based upon work supported by the Air Force Office of Scientific Research under award number FA9550-16-1-0362
titled Multidisciplinary Approaches to Radiation Balanced Lasers (MARBLE).
%%%%%%%%%%%%%%%%%%%%%%%%%%%%%%%%%%%%%%%%%%%%%%%%%%%%%%
%%%%%%%%%%%%%%%%%%%%%%%%%%%%%%%%%%%%%%%%%%%%%%%%%%%%%%
%\bibliography{cooling-refs.bib}

%%%%%%%%%%%%%%%%%%%%%%%%%%%%%%%%%%%%%%%%%%%%%%%%%%%%%%
%%%%%%%%%%%%%%%%%%%%%%%%%%%%%%%%%%%%%%%%%%%%%%%%%%%%%%
\end{document}